\documentclass[%
 aip,
 reprint,%
]{revtex4-1}
\usepackage{graphicx}
\usepackage{dcolumn}
\usepackage{bm}
\usepackage{amsmath,amssymb}

\usepackage[utf8]{inputenc}
\usepackage[T1]{fontenc}
\usepackage[libertine]{newtxmath}
\usepackage{libertineRoman}
\usepackage{etoolbox}

\makeatletter
\renewcommand*{\fnum@figure}{{\normalfont\bfseries\sffamily \figurename~\thefigure}}
\renewcommand*{\@caption@fignum@sep}{\textbf{ : }\sffamily}
\makeatother

\usepackage{algorithm}
\usepackage{algorithmicx}
\usepackage[noend]{algpseudocode}

\usepackage{xcolor}

\newcommand{\tr}{\operatorname{tr}}

\newcommand{\cT}{\dagger}

\newcommand{\PP}{\operatorname{Prob}}
\newcommand{\HA}{\operatorname{HA}}

\usepackage{bm}
\renewcommand{\vec}{}

\newcommand{\Id}{{I}}
\newcommand{\Tl}{{T}}
\newcommand{\Ql}{{Q}}

\newcommand{\nv}{n_{\textup{v}}}

\newcommand{\tot}{\textup{t}}
\newcommand{\sys}{\textup{s}}
\newcommand{\bath}{\textup{b}}
\newcommand{\syba}{\textup{sb}}

\usepackage{graphicx}
\usepackage{dsfont}

\usepackage{xcolor}

\usepackage{hyperref}
\usepackage[capitalize,nameinlink]{cleveref}

\usepackage[normalem]{ulem}

\usepackage{lipsum}

\makeatletter
\def\@email#1#2{%
 \endgroup
 \patchcmd{\titleblock@produce}
  {\frontmatter@RRAPformat}
  {\frontmatter@RRAPformat{\produce@RRAP{*#1\href{mailto:#2}{#2}}}\frontmatter@RRAPformat}
  {}{}
}%
\makeatother

\begin{document}

\preprint{AIP/123-QED}

\title[Numerical computation of the reduced density matrix]{Numerical computation of the equilibrium-reduced density matrix for \\strongly coupled open quantum systems}
\author{Tyler Chen}
\email{chentyl@uw.edu}
\homepage{https://chen.pw}
\affiliation{%
\mbox{Department of Applied Mathematics, University of Washington, Seattle, Washington 98195, USA}}

\author{Yu-Chen Cheng}
\email{yuchen@ds.dfci.harvard.edu}
\affiliation{%
\mbox{Department of Data Science, Dana-Farber Cancer Institute, Boston, Massachusetts 02215, USA}}
\affiliation{%
\mbox{Department of Biostatistics, Harvard T.H. Chan School of Public Health, Boston, Massachusetts 02115, USA}}
\affiliation{%
\mbox{Center for Cancer Evolution, Dana-Farber Cancer Institute, Boston, Massachusetts 02215, USA}}

\date{\today}

\begin{abstract}
We describe a numerical algorithm for approximating the equilibrium-reduced density matrix and the effective (mean force) Hamiltonian for a set of system spins coupled strongly to a set of bath spins when the total system (system+bath) is held in canonical thermal equilibrium by weak coupling with a ``super-bath''.
Our approach is a generalization of now standard typicality algorithms for computing the quantum expectation value of observables of bare quantum systems via trace estimators and Krylov subspace methods.
In particular, our algorithm makes use of the fact that the reduced system density, when the bath is measured in a given random state, tends to concentrate about the corresponding thermodynamic averaged reduced system density.
Theoretical error analysis and numerical experiments are given to validate the accuracy of our algorithm.
Further numerical experiments demonstrate the potential of our approach for applications including the study of quantum phase transitions and entanglement entropy for long-range interaction systems.
\end{abstract}

\maketitle

\section{Introduction}

The equilibrium thermodynamics of quantum systems is a growing area of research  \cite{gemmer_michel_mahler_09,vinjanampathy_anders_16,alicki_kosloff_18}, and in many cases, systems of interest are open; for instance because the system of interest is a subsystem of some closed system.
This presents a practical barrier to the study of open quantum systems, since open quantum systems enjoy many complexities not present in closed system.
Indeed, many thermodynamic properties of even the simplest open quantum systems may be seemingly anomalous; e.g. the specific heat or entropy may be negative \cite{ingold_hanggi_talkner_09,campisi_zueco_talkner_10,talkner_hanggi_20}.

In practice, only a relatively small number of quantum systems have Hamiltonians with known analytic diagonalizations, and while numerical techniques exist for certain systems, for instance a small system coupled to a harmonic reservoir \cite{makri_makarov_95,makri_makarov_95b,gelzinis_valkunas_20,trushechkin_merkli_cresser_anders_22,chiu_strathearn_keeling_22}, 
efficient algorithms for the case when the total system is comprised entirely of spins are seemingly less available. 
This paper aims to address this gap by introducing and analyzing an algorithm for computing equilibrium thermodynamic properties of small open quantum spin systems coupled arbitrarily to baths consisting of a small to moderate number of spins.

Consider a total system Hamiltonian $\vec{H}_{\tot}$ of the form
\begin{equation}
    \label{eqn:total_Hamiltonian_decomp}
    \vec{H}_{\tot} = \vec{\bar H}_{\sys} + \vec{\bar H}_{\bath} + \vec{H}_{\syba}
\end{equation}
where $\vec{\bar H}_{\sys} = \vec{H}_{\sys} \otimes \Id_{\bath}$ corresponds to the  Hamiltonian of the bare system, $\vec{\bar H}_{\bath} = \Id_{\sys} \otimes \vec{H}_{\bath}$ corresponds to the Hamiltonian of the bare bath, and $\vec{H}_{\syba}$ an interaction term accounting for non-negligible interactions between the system and bath.

Throughout, we assume the total system (system+bath) is in a canonical equilibrium state due to weak contact with a ``super-bath'' which holds the total system at inverse Boltzmann temperature $\beta = (k_BT)^{-1}$.
The states of the total system/bare system/bare bath are then described by the density matrices
\begin{equation}
    \rho_{\tot/\sys/\bath}(\beta) = \frac{\exp[-\beta \vec{H}_{\tot/\sys/\bath}]}{Z_{\tot/\sys/\bath}(\beta)}
\end{equation}
where the partition functions are computed by
\begin{equation}
    Z_{\tot/\sys/\bath}(\beta) = \tr(\exp[-\beta \vec{H}_{\tot/\sys/\bath}]).
\end{equation}
Due to system-bath entanglement, the density matrix $\vec{\rho}_{\sys}(\beta )$ for the bare system \emph{does not} describe the equilibrium state of the system when strongly coupled to the bath; i.e. when $H_{\syba} \neq \vec{\mathit{0}}_{\tot}$ \cite{talkner_hanggi_20}.
Instead, one must start with the total system density and ``trace out'' the effects of the bath. 
The resulting reduced system density matrix, sometimes called the mean force Gibbs state,  is given by
\begin{equation}
    \vec{\rho}^*(\beta) =  \tr_{\bath}(\vec{\rho}_{\tot}(\beta)).
\end{equation}
where $\tr_{\bath}(\:\cdot\:)$ is the partial trace with respect to the bath.

The reduced system density matrix $\vec{\rho}^*(\beta)$ can be expressed in terms of an effective Hamiltonian $\vec{H}^*(\beta)$, often called the Hamiltonian of the mean force, by the relation
\begin{equation} 
    \label{eqn:density_from_H}
    \vec{\rho}^*(\beta) = \frac{ \exp[-\beta \vec{H}^*(\beta) ]}{Z^*(\beta)}
\end{equation}
where the corresponding partition function is
\begin{equation}
    Z^*(\beta) = \tr(\exp[-\beta \vec{H}^*(\beta)])
    = \frac{Z_{\tot}(\beta)}{Z_{\bath}(\beta)}.
\end{equation}
Thus, the Hamiltonian of the mean force has an explicit formula
\begin{equation}
    \label{eqn:HMF}
    \vec{H}^*(\beta) 
    = -\frac{1}{\beta} \ln \left[ \frac{ \tr_{\bath}(\exp[ -\beta \vec{H}_{\tot}])}{Z_{\bath}(\beta)} \right].
\end{equation}
We turn readers to \cite{talkner_hanggi_20} for a more detailed discussion on the Hamiltonian of the mean force and mean force Gibbs state.

The partition function $Z^*(\beta)$ for the reduced system provides access to thermodynamic quantities of the reduced system. 
These quantities include the heat capacity, magnetization, susceptibility, etc. and are often functions of the Helmholtz free energy $F^*(\beta) = -\beta^{-1} \ln(Z^*(\beta))$ and can therefore be written in terms of the difference of the corresponding quantities of the bare system and bare bath.
In fact, since $Z^*(\beta)$ only depends on $Z_{\tot}(\beta)$ and $Z_{\bath}(\beta)$, any quantities depending on $Z^*(\beta)$ can be computed using what have become standard numerical techniques; see \cref{sec:past_work}.

Other properties of the reduced system cannot alone be derived from the partition functions of the bare system and bath.
Indeed, $Z_{\tot}(\beta)$ and $Z_{\bath}(\beta)$ do not account for the structure of interactions between the system and bath and therefore cannot express any quantities, such as the von Neumann entropy \cite{vonneumann_55}, which may depend on entanglement between the system and bath.
In these cases, one must obtain the Hamiltonian of the mean force $\vec{H}^*(\beta)$ or the corresponding density matrix $\vec{\rho}^*(\beta)$.
Computing such quantities numerically is the main focus of this paper.

\subsection{Notation}

We denote by $\mathcal{H}_{\sys}$ and $\mathcal{H}_{\bath}$ the Hilbert spaces for the system and bath so that $\mathcal{H}_{\tot} = \mathcal{H}_{\sys}\otimes \mathcal{H}_{\bath}$ is the Hilbert space for the total system.
For a Hilbert space $\mathcal{H}$, we denote by $|\mathcal{H}|$ the dimension of the space and by $\operatorname{L}(\mathcal{H})$ the set of self-adjoint linear operators on $\mathcal{H}$.
Throughout $| \kern 1pt \vec{i}\kern 1pt \rangle$ is the $i$-th standard basis vector of dimension determined by context and $\Id_{\tot/\sys/\bath}$ and $\vec{\mathit{0}}_{\tot/\sys/\bath}$ are identity and zero operators on $\mathcal{H}_{\tot/\sys/\bath}$.

\section{Background}

\subsection{Typicality}

Broadly, quantum typicality refers to the idea that, in many cases, a random state is representative of the overall state of a system.
Early concepts of typicality were hinted at by Schr\"odinger \cite{schrodinger_27} and proved rigorously by von Neumann \cite{vonneumann_29}; see \cite{goldstein_lebowitz_mastrodonato_tumulka_zanghi_10} for an overview.
Mathematically, the notion of typicality can be viewed as \emph{concentration} of a random variable about it's expectation value. 

Before we describe the notion of typicality on which our algorithm is based, we introduce a similar, yet mathematically equivalent, form of typicality which asserts that the quantum expectation value $\langle \vec{v} | \vec{O} | \vec{v} \rangle$ of an observable $\vec{O}$ in a state $|\vec{v}\rangle$ is overwhelmingly likely to be near to the quantum expectation value of $\vec{O}$, at least when $|\vec{v}\rangle$ is chosen randomly from a suitable ensemble. 

More precisely, suppose $\vec{O} \in\operatorname{L}(\mathcal{H}_{\tot})$ is an observable of the total system and $|\vec{v}\rangle\in\mathcal{H}_{\tot}$ is a random state sampled from the uniform distribution on the set of all states.
Then, the ``Hilbert space average" (HA) of the density matrix $|v\rangle \langle v|$ (with respect to the above distribution) is
\begin{align}
    \label{eqn:cov_assumption}
   \HA[|v\rangle \langle v|] =  |\mathcal{H}_{\tot}|^{-1}  \vec{I}_\tot.
\end{align}
Therefore, using basic properties of the trace and Hilbert space average, the quantum expectation value of $O$ when the system is in state $|\vec{v} \rangle$ satisfies 
\begin{align} 
    \HA \big[ \langle \vec{v} | 
    \vec{O} | \vec{v} \rangle \big] 
    &= \HA\big[ \tr(| \vec{v} \rangle \langle \vec{v} | \vec{O} )  \big] 
    \\&= \tr(\HA[ | \vec{v} \rangle \langle \vec{v} | ] \vec{O} ) 
    \\&= |\mathcal{H}_{\tot}|^{-1} \tr(\vec{O}).
\end{align}
Here $|\mathcal{H}_{\tot}|^{-1} \tr(\vec{O})$ is the quantum expectation of $\vec{O}$ when the state of the system is described by the density matrix $|\mathcal{H}_{\tot}|^{-1} \vec{I}_\tot$.

More generally, if the state of the system is described by an arbitrary density matrix $\vec{\rho}$, then the quantum expectation of an observable $\vec{O}$ is $\tr(\vec{\rho} \vec{O}) = \tr(\sqrt{\vec{\rho}}\vec{O}\sqrt{\vec{\rho}})$.
Therefore, in order to find an ensemble $|\vec{\omega}\rangle$ so that the quantum expectation value $\langle \vec{\omega} | \vec{O} | \vec{\omega} \rangle $ has Hilbert space average equal to $\tr(\vec{\rho} \vec{O})$, we simply define $|\vec{\omega}\rangle$ by $| \vec{\omega} \rangle := \sqrt{|\mathcal{H}_{\tot}|\rho} \,| \vec{v} \rangle$  where $|\vec{v}\rangle$ remains a uniformly chosen random state.
Early analyses \cite{reimann_07, bartsch2009dynamical, sugiura_shimizu_12} showed that  $\langle \vec{\omega} | O | \vec{\omega} \rangle $ concentrates about it's quantum expectation $\tr(\rho O)$ by bounding the variance of $\langle \vec{\omega} | O | \vec{\omega} \rangle$ and applying Chebyshev's inequality.
Subsequent analyses show that $\langle \vec{\omega} | O | \vec{\omega} \rangle $ is in fact sub-Gaussian and concentrates far more sharply about $\tr(\rho O)$ than suggested by Chebyshev's inequality.
Mathematically precise bounds are discussed in \cref{sec:error_analysis}.

We now introduce the version of typicality on which our algorithm is based.
Recall that the partial trace of $\vec{\rho} \vec{O}$ with respect to the bath Hilbert space can be expressed as
\begin{equation}
    \tr_{\bath}(\vec{\rho} \vec{O}) = \sum_{i=1}^{|\mathcal{H}_{\bath}|} ( \Id_{\sys} \otimes \langle\kern 1pt \vec{i}\kern 1pt| )   \vec{\rho} \vec{O}  (\Id_{\sys} \otimes | \kern 1pt \vec{i} \kern 1pt \rangle ).
\end{equation}
Thus, if $|\vec{v}\rangle\in\mathcal{H}_{\bath}$ is a random state chosen uniformly from all states we have $\HA[|\vec{v}\rangle\langle v|] = |\mathcal{H}_{\bath}|^{-1} \Id$, and it is relatively straightforward to see that
\begin{equation}
    \HA\big[ 
        (\Id_{\sys} \otimes \langle \vec{v} | )  \sqrt{|\mathcal{H}_{\bath}|\vec{\rho}}\, \vec{O} \sqrt{|\mathcal{H}_{\bath}|\vec{\rho}}\, (\Id_{\sys} \otimes | \vec{v} \rangle)
    \big] 
    = \tr_{\bath}(\vec{\rho} \vec{O}).
\end{equation}
Indeed, expand 
\begin{equation}
    |\vec{v}\rangle = \sum_{i=1}^{|\mathcal{H}_{\bath}|} \langle \kern 1pt \vec{i} \kern 1pt |\vec{v} \rangle | \kern 1pt \vec{i} \kern 1pt \rangle
\end{equation}
and observe that, for any $i,j=1,2,\ldots,|\mathcal{H}_{\bath}|$,
\begin{equation}
    \label{eqn:vivj}
    \HA\big[\langle \kern 1pt \vec{i} \kern 1pt |\vec{v} \rangle\langle \kern 1pt \vec{j} \kern 1pt |\vec{v} \rangle \big]
    = \begin{cases} 
    |\mathcal{H}_{\bath}|^{-1} & i=j
    \\
    0  & i \neq j
    \end{cases}.
\end{equation}
Then, using the linearity of the Hilbert space average and \eqref{eqn:vivj}, we find that, for any $\vec{A}\in\operatorname{L}(\mathcal{H}_{\tot})$,
\begin{align}
    &\HA\big[(\Id_{\sys} \otimes \langle \vec{v} | ) \vec{A} (\Id_{\sys} \otimes | \vec{v} \rangle )\big]
    \\&\hspace{2em}=
    \sum_{i=1}^{|\mathcal{H}_\tot|}\sum_{j=1}^{|\mathcal{H}_\tot|}
    \HA\big[\langle \kern 1pt \vec{i} \kern 1pt |\vec{v} \rangle\langle \kern 1pt \vec{j} \kern 1pt |\vec{v} \rangle \big]
    (\Id_{\sys} \otimes \langle \kern 1pt \vec{i} \kern 1pt | )  \vec{A} (\Id_{\sys} \otimes | \kern 1pt \vec{j} \kern 1pt \rangle )
    \\&\hspace{2em}=
    \sum_{i=1}^{|\mathcal{H}_{\bath}|} |\mathcal{H}_{\bath}|^{-1}( \Id_{\sys} \otimes \langle\kern 1pt \vec{i}\kern 1pt| )   \vec{A}  (\Id_{\sys} \otimes | \kern 1pt \vec{i} \kern 1pt \rangle )
    \\&\hspace{2em}=
    |\mathcal{H}_{\bath}|^{-1}\tr_{\bath}(\vec{A}).
\end{align}

In fact, a simple union bound shows that $( \Id_{\sys} \otimes \langle \vec{v} | ) \sqrt{|\mathcal{H}_{\bath}|\vec{\rho}}\, \vec{O} \sqrt{|\mathcal{H}_{\bath}|\vec{\rho}}\, (\Id_{\sys} \otimes | \vec{v} \rangle )$ also exhibits sub-Gaussian concentration about it's mean.
From a linear algebraic perspective, this is equivalent to previously defined versions of typicality. 
Even so, we were unable to find explicit reference to this phenomenon in the literature.

\subsection{Spin systems}

For concreteness, we will consider total systems consisting of $N$ interacting spins sites of spin number $s$ in a magnetic field of strength $h$ pointing in the $\textup{z}$-direction. 
The corresponding Heisenberg Hamiltonian for the total system is
\begin{equation}
\label{eqn:Ht}
    \vec{H}_{\tot} = 
    \sum_{i,j=1}^{N} {J}_{i,j} \cdot {\sigma}_i {\sigma}_j
    + \frac{h}{2} \sum_{i=1}^{N} \sigma_i^{\textup{z}}
\end{equation}
where we have used the shorthand
\begin{equation}
    {J}_{i,j} \cdot {\sigma}_i {\sigma_j}
    = J^{\textup{x}}_{i,j} \vec{\sigma}^{\textup{x}}_i \vec{\sigma}^{\textup{x}}_j 
    +J^{\textup{y}}_{i,j} \vec{\sigma}^{\textup{y}}_i \vec{\sigma}^{\textup{y}}_j
    +J^{\textup{z}}_{i,j} \vec{\sigma}^{\textup{z}}_i \vec{\sigma}^{\textup{z}}_j,
\end{equation}
where $J_{i,j}^{\textup{x}/\textup{y}/\textup{z}}$ describes the coupling strength between sites $i$ and $j$ in the ${\textup{x}/\textup{y}/\textup{z}}$ coordinate directions.
Here \( \vec{\sigma}_i^{\textup{x}/\textup{y}/\textup{z}} \) gives the component spin operator for the \( i \)-th spin site and acts trivially on the Hilbert spaces associated with other spin sites but as the \( (2s+1)\times (2s+1) \) component spin matrix \( \vec{\sigma}^{\textup{x}/\textup{y}/\textup{z}} \) on the \( i \)-th spin site.
In matrix form, \( \vec{\sigma}_i^{\textup{x}/\textup{y}/\textup{z}} \) is written
\begin{equation}
    \vec{\sigma}_i^{\textup{x}/\textup{y}/\textup{z}}
    = \underbrace{\Id \otimes \cdots \otimes \Id}_{i-1\text{ terms}} 
    \otimes ~ \vec{\sigma}^{\textup{x}/\textup{y}/\textup{z}} \otimes 
    \underbrace{\Id \otimes \cdots \otimes \Id}_{N-(i-1)\text{ terms}}.
\end{equation}

Without loss of generality, we will take the system to be the first $N_{\sys}$ sites and the bath to be the remaining $N_{\bath} = N - N_{\sys}$ sites.
Let $\mathcal{I}_{\sys} = \{ 1,\ldots, N_{\sys} \}$ and $\mathcal{I}_{\bath} = \{N_{\sys}+1,\ldots, N \}$.
$\vec{H}_{\tot}$ can then be decomposed according to \eqref{eqn:total_Hamiltonian_decomp} where
\begin{align}
    \vec{\bar{H}}_{\sys} &= \sum_{i,j \in \mathcal{I}_\sys} {J}_{i,j} \cdot {\sigma}_i {\sigma}_j
    + \frac{h}{2} \sum_{i \in \mathcal{I}_{\sys}} \vec{\sigma}_i^{\textup{z}}
\\
    \vec{\bar{H}}_{\bath} &= \sum_{i,j \in \mathcal{I}_{\bath}} {J}_{i,j} \cdot {\sigma}_i {\sigma}_j
    + \frac{h}{2} \sum_{i \in \mathcal{I}_{\bath}} \vec{\sigma}_i^{\textup{z}}
    \\
    \vec{H}_{\syba} &= \sum_{\substack{i\in \mathcal{I}_{\sys},j\in \mathcal{I}_{\bath}\\j\in \mathcal{I}_{\sys},i\in \mathcal{I}_{\bath}}} {J}_{i,j} \cdot {\sigma}_i {\sigma}_j.
\end{align}

\subsection{High and low temperature limits}
\label{sec:limits}

From the structure of the total system Hamiltonian \eqref{eqn:Ht} considered in this paper, we can derive the limits for the mean force Hamiltonian and mean force Gibbs state at high and low temperature.
We summarize these limits here, and provide proofs in \cref{sec:Hmf_limit}.

In the high temperature limit, as $\beta \to 0$, it is straightforward, although a bit tedious, to verify that $\vec{H}^*(\beta) = \vec{H}_{\sys} + \tr_{\bath}(\vec{H}_{\syba}) / \tr(\Id_{\bath}) + \mathcal{O}(\beta)$. 
Since $\vec{H}_{\syba}$ accounts for interactions between the system and bath (and therefore contains no interactions between spin sites within the bath) we have that $\tr_{\bath}(\vec{H}_{\syba}) = \vec{\mathit{0}}_{\sys}$.
This implies that,
\begin{equation}
    \lim_{\beta\to 0} \vec{H}^*(\beta) 
    = \vec{H}_{\sys}.
\end{equation}
In this case, $\vec{\rho}^*(\beta) \to (2s+1)^{-N_{\sys}} \Id_{\sys}$.


In the low temperature limit, as $\beta\to\infty$, we have that $\vec{\rho}_{\tot}(\beta)$ is convergent to $r^{-1} \sum_{i=1}^{r} | \vec{\psi}_i \rangle \langle \vec{\psi}_i |$, where $\{| \vec{\psi}_i \rangle\}$ are the $r$ ground states for the total system 
This implies that $\vec{\rho}^*(\beta)$ is also convergent to a fixed density matrix.
Specifically, 
\begin{equation}
    \lim_{\beta\to\infty} \vec{\rho}^*(\beta) 
    = \frac{1}{r} \sum_{i=1}^{r} \tr_{\bath}( | \vec{\psi}_i \rangle \langle \vec{\psi}_i | ).
\end{equation}
Then, since $\vec{H}^*(\beta) = -\beta^{-1} \ln[Z^*(\beta)\vec{\rho}^*(\beta)]$ it is easy to verify that
\begin{equation}
    \lim_{\beta\to\infty} \vec{H}^*(\beta) 
    = \lim_{\beta\to\infty} \ln(Z^*(\beta)) \Id
    = (E_{\tot} - E_{\bath}) \Id,
\end{equation}
where $E_{\tot}$ and $E_{\bath}$ are the ground state energies of $\vec{H}_{\tot}$ and $\vec{H}_{\bath}$ respectively.

\subsection{Past algorithms}
\label{sec:past_work}

A major challenge to the design of algorithms for quantum spin systems is that problem sizes grow exponentially with the total system size.
Perhaps the simplest numerical approach is to apply an exact eigensolver to $\vec{H}_{\tot}$ in order to compute $\exp(-\beta \vec{H}_{\tot})$. 
However, this quickly becomes intractable unless there are many symmetries present in the system.
Moreover, even if $\exp(-\beta \vec{H}_{\tot})$ could somehow be computed efficiently, the cost of storing it would be very high.
For instance, if the total system is comprised of $20$ spin-$\tfrac{1}{2}$ particles, $\exp(-\beta \vec{H}_{\tot})$ is a matrix with $2^{20}\times 2^{20}$ entries which would require nearly 8.8 \emph{terrabytes} of memory to store in a 64 bit double precision format \footnote{While $\vec{H}_{\tot}$ may be sparse, the matrix exponential is not, in general, sparse.}.

Over the past several decades, typicality based approaches, such as the finite temperature Lanczos method (FTLM), have become among the most widely used numerical methods for approximating equilibrium thermodynamic properties of closed quantum systems
\cite{skilling_89,jaklic_prelovsek_94,schnalle_schnack_10,kpm_review_06,schnack_richter_steinigeweg_20,schulter_gayk_schmidt_honecker_schnack_21,jin_willsch_willsch_lagemann_michielsen_deraedt_21}.
Such methods make use of Krylov subspace methods to avoid explicitly forming matrix functions such as $\exp(-\beta \vec{H}_{\tot})$, and instead compute quantities like $\langle \vec{v} | \exp(-\beta \vec{H}_{\tot})| \vec{v} \rangle$.
Moreover, because Krylov subspace methods are ``matrix-free'', they only access $\vec{H}_{\tot}$ through matrix-vector products.
Theoretical analyses of such algorithms is an active area of research \cite{han_malioutov_avron_shin_17,ubaru_chen_saad_17,chen_trogdon_ubaru_22}.

Krylov subspace methods can also be used to numerically compute the mean force Hamiltonian and reduced system density matrix at high and low temperature using the limits from \cref{sec:Hmf_limit}.
Indeed, at high temperature $\rho^*(\beta)$ is trivial and $\vec{H}^*(\beta)$ converges to $\vec{H}_{\sys}$.
At low temperature, $\rho^*(\beta)$ depends only on the ground state(s) of $\vec{H}_{\tot}$ and $\vec{H}^*(\beta)$ depends only on the ground state energies of $\vec{H}_{\tot}$ and $\vec{H}_{\bath}$.

We emphasize that the task of computing a partial trace of an \emph{explicit} matrix, even naively, is not particularly difficult.
More efficient approaches to this task have also been studied \cite{maziero_17}.
However, to the best of our knowledge, the task of computing the partial trace of \emph{implicit} matrices such as $\exp(-\beta \vec{H}_{\tot})$, without ever explicitly constructing the matrix, has not been thoroughly studied.

\section{Algorithm}

In this section, we describe a numerical method for computing the mean force Hamiltonian and reduced system density matrix.
In the case of an empty system, so that $\mathcal{H}_{\tot} = \mathcal{H}_{\bath}$, everything in this section reduces to the well-known stochastic Lanczos quadrature algorithm \cite{bai_fahey_golub_96,ubaru_chen_saad_17,chen_trogdon_ubaru_22}.
A theoretical error analysis is given in \cref{sec:error_analysis}.

\subsection{Block stochastic Lanczos quadrature}

Let $\vec{A} \in \operatorname{L}(\mathcal{H}_{\tot})$ and recall that $( \Id_{\sys} \otimes \langle \vec{v} | )  \vec{A} (\Id_{\sys} \otimes | \vec{v} \rangle )$ is an unbiased estimator for $|\mathcal{H}_{\bath}|^{-1}\tr_{\bath}(\vec{A})$.
To improve the estimator's accuracy, we can average multiple copies.
Specifically, given $\nv$ iid samples $\{ | \vec{v}_j \rangle \}$ of $| \vec{v} \rangle$ (with $\HA[|\vec{v}\rangle\langle \vec{v} |] = \vec{I}_{\bath}$, we can define the averaged estimator
\begin{equation}
    \label{eqn:trace_est}
    \frac{1}{\nv}\sum_{i=1}^{\nv} ( \Id_{\sys} \otimes \langle \vec{v}_j | )  \vec{A} (\Id_{\sys} \otimes | \vec{v}_j \rangle ).
\end{equation}
Physically, we can view this averaging technique as constructing a new total system containing multiple independent copies of the original total system. This point of view has its origins in Gibbs's 1902 concept of the statistical ensemble \cite{gibbs1902elementary}. 
While many analyses from physics \cite{goldstein_lebowitz_mastrodonato_tumulka_zanghi_10, bartsch2009dynamical, sugiura_shimizu_12}
show that a high dimensional Hilbert space $|\mathcal{H}_{\bath}|$ can lead to quantum typicality under regular conditions given in their models, in numerical analysis, it is more common to consider the error of estimator \eqref{eqn:trace_est} when $\nv$ is large.
We believe the mathematical correspondence between $|\mathcal{H}_{\bath}|$ and $\nv$ is worthy of a more rigorous mathematical analysis.
For instance, generalized fundamental thermodynamic relations were recently unraveled by replacing thermodynamic infinite-size limit ($|\mathcal{H}_{\bath}| \rightarrow \infty$) with multiple-measurement limit ($\nv \rightarrow \infty$)  \cite{lu2022emergence}. 

Often $\vec{A} = f[\vec{H}]$ for some function $f:\mathbb{R}\to\mathbb{R}$ and $\vec{H}\in \operatorname{L}(\mathcal{H}_{\tot})$; for instance $f = x\mapsto \exp(-\beta x)$ and $\vec{H}=\vec{H}_{\tot}$.
If $\vec{H}$ has known diagonalization, then we can easily compute $( \Id_{\sys} \otimes \langle \vec{v} | ) f[\vec{H}] (\Id_{\sys} \otimes | \vec{v} \rangle )$.
Unfortunately, diagonalizing large Hermitian matrices is often exceedingly expensive.
In fact, even for highly structured matrices, such as those considered in this paper, exact diagonalization may be too costly.

A natural approach to avoiding such costs is to apply the block Lanczos algorithm \cref{alg:block_lanczos} to $\vec{H}$ and $\Id_{\sys} \otimes | \vec{v} \rangle$ for $k$ iterations to obtain a $k|\mathcal{H}_{\sys}| \times k|\mathcal{H}_{\sys}|$ block tridiagonal matrix $\Tl$.
Given a matrix $\vec{V}$ with $\vec{V}^\cT \vec{V} = \Id_{\sys}$, \cref{alg:block_lanczos} computes orthonormal matrices $\{\vec{Q}_j\}$, $j=1, \ldots, k+1$ such that $\vec{Q}_i^\cT \vec{Q}_j = \delta_{i,j} \Id_{\sys}$ and for all $j\leq k$,
\begin{equation}
\operatorname{span}\{ \vec{V}, \vec{H}\vec{V}, \ldots, \vec{H}^j \vec{V} \} = \operatorname{span}\{ \vec{Q}_1, \vec{Q}_2, \ldots, \vec{Q}_j \}.
\end{equation}
These vectors satisfy a symmetric block-tridiagonal recurrence
\begin{equation}
\label{eqn:krylov_recurrence}
    \vec{H} \Ql = \Ql \Tl + \vec{Q}_{k+1} \vec{B}_k \vec{E}_k^\cT
\end{equation}
where $\vec{E}_k = | \kern 1pt \vec{k} \kern 1pt \rangle \otimes \Id_{\sys}$ and
\begin{equation}
    \Tl = \begin{bmatrix}
    \vec{A}_1 & \vec{B}_1^\cT \\
    \vec{B}_1 & \ddots & \ddots \\
    & \ddots & \ddots & \vec{B}_{k-1}^\cT \\
    &&\vec{B}_{k-1} & \vec{A}_k
    \end{bmatrix}
    ,~~
    \Ql = \begin{bmatrix}
    | &|&  & |\\
    \vec{Q}_1 & \vec{Q}_2 & \cdots & \vec{Q}_k \\
    | &|&  & |
\end{bmatrix}.
\end{equation}

\begin{algorithm}[H]
\caption{Block Lanczos}
\label{alg:block_lanczos}
\fontsize{10}{10}\selectfont
\begin{algorithmic}[1]
\Procedure{block-Lanczos}{$\vec{H}, \vec{V}, k$}
\State \( \vec{Q}_1 = \vec{V} \),
\For {\( j=1,2,\ldots,k \)}
    \State \( \vec{Z} = \vec{H} \vec{Q}_{j} - \vec{Q}_{j-1} \vec{B}_{j-1}^\cT \)
    \State \( \vec{A}_j = \vec{Q}_{j}^\cT \vec{Z}  \)
    \State \( \vec{Z} = \vec{Z} - \vec{Q}_{j} \vec{A}_j \)
    \State \( \vec{Q}_{j+1},\vec{B}_{j} = \textsc{qr}(\vec{Z}) \)
\EndFor
\State \Return $\{\vec{Q}_j\}, \{\vec{A}_j\}, \{ \vec{B}_j \}$
\EndProcedure
\end{algorithmic}
\end{algorithm}

The expression $( \Id_{\sys} \otimes \langle \vec{v} | ) f[\vec{H}] (\Id_{\sys} \otimes | \vec{v} \rangle )$ can be approximated by 
\begin{equation}
    \label{eqn:fT}
    (\langle  \vec{\mathit{1}} | \otimes \Id_{\sys}) f[\Tl] ( | \vec{\mathit{1}} \rangle \otimes \Id_{\sys}).
\end{equation}
Assuming $k|\mathcal{H}_{\sys}|$ is small enough so that $\Tl$ can be diagonalized exactly, then $f[\Tl]$ can be computed directly.
This is a \emph{block Gauss quadrature} approximation and is exact if $f$ is a polynomial of degree at most $2k-1$ \cite[Section 6.6]{golub_meurant_09}.

To obtain our final estimator, we apply this approach to each term in \eqref{eqn:trace_est}. 
Specifically, for $j=1,\ldots, \nv$ denoting by $\Tl_j$ the resulting block-tridiagonal matrix obtained by block Lanczos run on $\vec{H}$ and $\Id_{\sys}\otimes | \vec{v}_j \rangle$, our estimator for $\tr_{\bath}(f[\vec{H}])$ is
\begin{equation}
    \label{eqn:trB_est}
    \frac{1}{\nv} \sum_{j=1}^{\nv} (\langle  \vec{\mathit{1}} | \otimes \Id_{\sys}) f[\Tl_j] ( | \vec{\mathit{1}} \rangle \otimes \Id_{\sys}).
\end{equation}

\subsubsection{Costs}

The accuracy and computational cost of \eqref{eqn:trB_est} depend on both $\nv$ and $k$.
Application of the block Lanczos method to each term of $\eqref{eqn:trB_est}$ uses $k$ block matrix-vector products with blocks of $|\mathcal{H}_{\sys}|$ vectors along with $\mathcal{O}(|\mathcal{H}_{\tot}||\mathcal{H}_{\sys}|) = \mathcal{O}(|\mathcal{H}_{\bath}||\mathcal{H}_{\sys}|^2)$ storage.
While a block matrix-vector product requires $|\mathcal{H}_{\sys}|$ times as many operations as a standard inner product, in practice block matrix-vector products can often be computed nearly as quickly as a single matrix-vector product.
If each term in \eqref{eqn:trB_est} is computed sequentially, the computational cost of computing \eqref{eqn:trB_est} scales linearly with $\nv$ while the storage cost is constant.
On the other hand, all terms can be computed entirely in parallel at the cost of $\nv$ times more storage.

Like other typicality algorithms, our approach requires matrix-products with the total system Hamiltonian $H_{\tot}$ and is therefore limited to systems for which it is possible to store several vectors of size $|\mathcal{H}_{\tot}|$.
Our algorithm is also limited in terms of the system size, since we must store a number of vectors of length $|\mathcal{H}_{\tot}|$ proportional to the system dimension $|\mathcal{H}_{\sys}|$.
Even so, our algorithm is significantly more efficient than exact diagonalization based techniques.

\subsection{An algorithm for the mean force Hamiltonian}

To compute $\vec{H}^*(\beta)$ we can use the same test states $\{ | \vec{v}_j \rangle\}$ for estimating the partial trace and the trace over the bath. 
Specifically, if $(\Tl_{\tot})_j$ is the block-tridiagonal matrix produced by the block Lanczos algorithm with $\vec{H}_{\tot}$ and $(\Id_{\sys} \otimes | \vec{v}_j \rangle )$ and $(\Tl_{\bath})_j$ is the tridiagonal matrix produced by Lanczos with $\vec{H}_{\bath}$ and $| \vec{v}_j \rangle$, by \eqref{eqn:HMF} we obtain the estimators
\begin{align}
    \vec{H}^*(\beta) 
    &\approx
     -\frac{1}{\beta} \ln \left[ \frac{\sum_{j=1}^{\nv} ( \Id_{\sys} \otimes \langle \vec{v}_j |) \exp[-\beta \vec{H}_{\tot}] ( \Id_{\sys}\otimes | \vec{v}_j \rangle) }{\sum_{j=1}^{\nv} \langle \vec{v}_j | \exp[-\beta \vec{H}_{\bath}] | \vec{v}_j \rangle  } \right]
    \label{eqn:H_est_exact}
    \\&\approx
    -\frac{1}{\beta} \ln \left[ \frac{\sum_{j=1}^{\nv} (\langle \vec{\mathit{1}} | \otimes \Id_{\sys}) \exp[-\beta (\Tl_{\tot})_j] ( | \vec{\mathit{1}} \rangle \otimes \Id_{\sys}) }{\sum_{j=1}^{\nv} \langle \vec{\mathit{1}} | \exp[-\beta (\Tl_{\bath})_j] | \vec{\mathit{1}} \rangle  } \right].
    \label{eqn:H_est}
\end{align}
Of course, only the latter estimator is practically computable.

When $\beta$ is large and the eigenvalues of $(\Tl_{\tot})_j$ or $(\Tl_{\bath})_j$ are negative, then the computation of the exponential may overflow. 
To avoid this, we can simply replace $x\mapsto \exp(-\beta x)$ with $x\mapsto \exp(-\beta(x-E_0))$ where $E_0$ chosen to be smaller than the smallest eigenvalues of $(\Tl_{\tot})_j$ and $(\Tl_{\bath})_j$; for instance, chosen to be (an estimate of) the smallest eigenvalue of $\vec{H}_{\tot}$.

It is worth noting that if we use \eqref{eqn:H_est} or \eqref{eqn:H_est_exact} to compute an approximation to $\vec{\rho}^*(\beta)$, by \eqref{eqn:density_from_H} we obtain exactly the estimators
\begin{align}
    \label{eqn:rho_est}
    \vec{\rho}^*(\beta) 
    &\approx
    \frac{\sum_{j=1}^{\nv} ( \Id_{\sys} \otimes \langle \vec{v}_j |) \exp[-\beta \vec{H}_{\tot}] ( \Id_{\sys}\otimes | \vec{v}_j \rangle) }{\tr\!\big( \sum_{j=1}^{\nv} (\langle \vec{v}_j | \otimes \Id_{\sys}) \exp[-\beta \vec{H}_{\tot}] ( | \vec{v}_j \rangle \otimes \Id_{\sys}) \big)}
    \\&\approx
    \frac{\sum_{j=1}^{\nv} (\langle \vec{\mathit{1}} | \otimes \Id_{\sys}) \exp[-\beta (\Tl_{\tot})_j] ( |\vec{\mathit{1}} \rangle \otimes \Id_{\sys}) }{\tr\!\big( \sum_{j=1}^{\nv} (\langle \vec{\mathit{1}} | \otimes \Id_{\sys}) \exp[-\beta (\Tl_{\tot})_j] ( | \vec{\mathit{1}} \rangle \otimes \Id_{\sys}) \big)}.
\end{align}

Thus, if one requires the eigenvalues of $\vec{H}^*(\beta)$ and $\rho^*(\beta)$, we suggest first computing the eigenvalues of $\sum_{j=1}^{\nv} (\langle \vec{\mathit{1}} | \otimes \Id_{\sys}) \exp[-\beta (\Tl_{\tot})_j] ( |\vec{\mathit{1}} \rangle \otimes \Id_{\sys})$, and then transforming them to obtain the eigenvalues of $\vec{H}^*(\beta)$ and $\rho^*(\beta)$.
This avoids the need for the computation of the matrix logarithm.

\section{Error analysis}
\label{sec:error_analysis}

In this section, we discuss bounds which provide intuition on how to balance $\nv$ and $k$.
For the block-size one case, such bounds have been studied extensively; see \cite{chen_trogdon_ubaru_22} for a recent review.

\subsection{Trace estimators}

Tail bounds trace estimators were studied in \cite{avron_toledo_11,roostakhorasani_ascher_14} with more recent and refined analyses are given in \cite{cortinovis_kressner_21,meyer_musco_musco_woodruff_21,chen_trogdon_ubaru_21}; see \cite{chen_trogdon_ubaru_22} for historical context.
For constants $\epsilon>0$ and $\delta\in(0,1)$, and $\vec{B} \in \operatorname{L}(\mathcal{H}_{\bath})$, these analyses aim to bound the number of samples $\nv$ required so that 
\begin{equation}
    \PP\bigg[ \Big| \frac{\tr(\vec{B})}{|\mathcal{H}_{\bath}|}  - \frac{1}{\nv} \sum_{j=1}^{\nv} \langle \vec{v}_j | \vec{B} | \vec{v}_j \rangle \Big| > \epsilon \bigg] < \delta.
\end{equation}
The resulting bounds are typically simple functions which depends on the distribution of $|\vec{v} \rangle$, the value of $\nv$, and basic properties of $\vec{B}$ such as its operator norm, Frobenius norm, or dimension.
Roughly speaking, the analyses in \cite{persson_cortinovis_kressner_22,chen_trogdon_ubaru_22} imply that, for small $\epsilon$, it suffices to take $\nv = \mathcal{O}( \| \vec{B} \|_2^2  \| \mathcal{H}_{\bath} |^{-1} \epsilon^2  ) \ln(2/\delta)$. Here $\tilde{\mathcal{O}}$ is equivalent to $\mathcal{O}$, but with poly-logarithmic factors in the constituent parameters suppressed for readability; i.e. we say a variable is $\tilde{\mathcal{O}}(h(t))$ if, for some $k\geq 0$, the variable is $\mathcal{O}(\log(t)^k h(t))$.

We can leverage such results to provide similar bounds for our partial trace estimator.
Towards this end, decompose $\vec{A}\in\operatorname{L}(\mathcal{H}_{\tot})$ as
\begin{equation}
    \vec{A} = \sum_{m,n=1}^{|\mathcal{H}_{\sys}|}  | \vec{m}  \rangle \langle \vec{n} |\otimes \vec{A}_{m,n}  
\end{equation}
where $\vec{A}_{m,n} \in \operatorname{L}(\mathcal{H}_{\bath})$. 
Fix $\epsilon >0$ and $\delta \in (0,1)$, and suppose that for all $m,n$ we have chosen $\nv$ so that
\begin{equation}
    \PP \Bigg[ \bigg| \frac{\tr(\vec{A}_{m,n})}{|\mathcal{H}_{\bath}|} - \frac{1}{\nv} \sum_{j=1}^{\nv}  \langle \vec{v}_j | \vec{A}_{m,n} | \vec{v}_j \rangle \bigg| > \tilde \epsilon \Bigg] < \tilde \delta
\end{equation}
for some $\tilde \epsilon$ and $\tilde\delta$, the exact values of which will soon become apparent.
Applying a union bound over all pairs $m,n$ we obtain the bound
\begin{equation}
    \PP \Bigg[ \forall m,n : \bigg|  \frac{\tr(\vec{A}_{m,n})}{|\mathcal{H}_{\bath}|} - \frac{1}{\nv} \sum_{j=1}^{\nv}  \langle \vec{v}_j | \vec{A}_{m,n} | \vec{v}_j \rangle  \bigg| > \tilde \epsilon \Bigg] < | \mathcal{H}_{\sys} |^2 \tilde \delta.
\end{equation}
Next, note that 
\begin{equation}
    \tr_{\bath}(\vec{A}) =  \sum_{m,n=1}^{|\mathcal{H}_{\sys}|}  \tr(\vec{A}_{m,n}) |\vec{m}  \rangle \langle \vec{n} |
\end{equation}
and that  
\begin{align}
    &\langle \vec{m} |  ( \Id_{\sys} \otimes \langle \vec{v}_j | )  \vec{A} (\Id_{\sys} \otimes | \vec{v}_j \rangle ) | \vec{n} \rangle 
    \nonumber \\&\hspace{4em}= ( \langle \vec{m} | \otimes \langle \vec{v}_j | )  \vec{A} ( | \vec{n} \rangle \otimes | \vec{v}_j \rangle )
    = \langle \vec{v}_j | \vec{A}_{m,n} | \vec{v}_j \rangle.
\end{align}
For any $\vec{X}\in\operatorname{L}(\mathcal{H}_{\sys})$ we have that $\|\vec{X}\| \leq | \mathcal{H}_{\sys} | \max_{m,n} \langle \vec{m}|\vec{X}|\vec{n} \rangle$.
Putting everything together we find that
\begin{equation}
    \PP \Bigg[ \bigg\|  \frac{\tr_{\bath}(\vec{A})}{|\mathcal{H}_{\bath}|} - \frac{1}{\nv} \sum_{j=1}^{\nv} ( \Id_{\sys} \otimes \langle \vec{v}_j| )  \vec{A} (\Id_{\sys} \otimes | \vec{v}_j \rangle )  \bigg\| > \epsilon \Bigg] <  \delta
\end{equation}
if we take $\tilde \epsilon = \epsilon / |\mathcal{H}_{\sys}| $ and $\tilde \delta = \delta / | \mathcal{H}_{\sys} |^2 $.
This allows existing bounds for standard trace estimators to be easily carried over to partial trace estimation. 
Like the basic trace estimators, since \eqref{eqn:trace_est} is the average of $\nv$ iid samples, $\nv$ must scale like $\mathcal{O}(\epsilon^2)$.

\subsubsection{A note on high temperatures}

At high temperature, our estimator will compute $\tr_{\bath}(\exp(-\beta \vec{H}_{\tot}))$ efficiently using only a single sample. 
This is because the randomness in our sample state is averaged over many states of the system, thereby reducing the variance of the output.

This does not necessarily imply an accurate estimate to $\vec{H}^*(\beta)$ using a single sample.
Indeed, recall that $\vec{H}^*(\beta) = \vec{H}_{\sys} + \tr_{\bath}(\vec{H}_{\syba}) / \tr(\Id_{\bath}) + \mathcal{O}(\beta)$. 
Thus, at high temperature, the approximation \eqref{eqn:H_est} differs from $\vec{H}_{\sys}$ by an additive factor 
$n_\textup{v}^{-1}\sum_{j=1}^{\nv}(\langle \vec{v}_j | \otimes \Id_{\sys}) \vec{H}_{\syba} (|\vec{v}_j \rangle \otimes \Id_{\sys}) / \tr(\Id_{\bath})$ which should be corrected for.
Since $\HA[(\langle \vec{v}_j | \otimes \Id_{\sys}) \vec{H}_{\syba}  (|\vec{v}_j \rangle \otimes \Id_{\sys})] = \vec{\mathit{0}}_{\sys}$, it may be possible to use this difference as a rough indicator of the accuracy of the averaged partial trace estimator.

\subsubsection{A note on low temperatures}

At low temperature, $\exp(-\beta\vec{H}_{\tot})$ is dominated by the ground state, or a few states near the ground state. 
As such, our approach will have relatively high variance because the randomness in our sample state is averaged over only several states. 
Thus, while the estimator still provides an unbiased estimate for $\tr_{\bath}(\exp(-\beta \vec{H}_{\tot}))$, many samples are required. 
In this setting, other approaches such as low rank approximation or hybrid methods make more sense \cite{lin_16,gambhir_stathopoulos_orginos_17,meyer_musco_musco_woodruff_21,persson_cortinovis_kressner_22,chen_hallman_22}.

\subsection{Block Gauss quadrature}

Recall that $\vec{V}^\cT p(\vec{A}) \vec{V} = \vec{E}_1^\cT p[\Tl] \vec{E}_1$ for all polynomials $p$ of degree at most $2k-1$.
For brevity, let $\textsc{Error} = \| \vec{V}^\cT f[\vec{H}] \vec{V} - \vec{E}_1^\cT f[\Tl] \vec{E}_1 \|$.
Then, for any polynomial $p$ with $\deg(p) \leq 2k-1$,
\begin{align}
    \hspace{1em}&\hspace{-1em}
    \textsc{Error} \nonumber
    \\&= \| \vec{V}^\cT f[\vec{H}] \vec{V} -\vec{V}^\cT p(\vec{A}) \vec{V} + \vec{E}_1^\cT p[\Tl] \vec{E}_1 - \vec{E}_1^\cT f[\Tl] \vec{E}_1 \| 
    \\&\leq \| \vec{V}^\cT (f[\vec{H}] - p[\vec{H}]) \vec{V}   \| +  \|  \vec{E}_1^\cT ( p[\Tl] - f[\Tl] ) \vec{E}_1 \| 
    \\ &\leq \| f[\vec{H}] - p[\vec{H}] \|  +  \|  p[\Tl] - f[\Tl] \|
    \\ & \leq 2\max_{x\in[\lambda_{\textup{min}},\lambda_{\textup{max}}]} | f(x) - p(x) |.
\end{align}
Here we have used that $\| \vec{V} \| \leq 1$ since $\vec{V}^\cT \vec{V} = \vec{I}$.
Optimizing over $p$ with $\deg(p) \leq 2k-1$ we find that
\begin{equation}
    \textsc{Error}
    \leq 2 \min_{\deg(p)\leq 2k-1} \bigg(\max_{x\in[\lambda_{\textup{min}},\lambda_{\textup{max}}]} | f(x) - p(x) | \bigg).
\end{equation}
Note that while we could instead apply the regular Lanczos algorithm to each of the vectors in $\vec{V}$ individually, the resulting algorithm would only be exact for polynomials of degree at most $k-1$.

Analytic functions such as the exponential can be approximated by polynomials of degree growing just logarithmically with the desired accuracy \cite{trefethen_19}.
This means that $k$ typically does not need to be very large.
Then, as long as $|\mathcal{H}_{\sys}|$ is also not too large, we can directly diagonalize each $\Tl_j$ to compute terms of \eqref{eqn:trB_est}, possibly exploiting the block tridiagonal structure along the way.

\subsubsection{Finite precision arithmetic}

In finite precision arithmetic, the output of the (block) Lanczos algorithm may be \emph{significantly} different than what would be obtained in exact arithmetic. 
In particular, the columns of $\vec{Q}$ may lose orthogonality. 
This has lead to some hesitance to use Lanczos based approaches without using costly explicit reorthogonalization \cite{jaklic_prelovsek_94,aichhorn_daghofer_evertz_vondelinden_03,kpm_review_06,ubaru_chen_saad_17}.

Careful analysis of the Lanczos algorithm in finite precision arithmetic \cite{paige_76,paige_80} can be leveraged to show that the Lanczos algorithm still works well for the task of applying matrix-functions to vectors  \cite{musco_musco_sidford_18} and quadratic forms \cite{knizhnerman_96} even in finite precision arithmetic.
In effect, these analyses show that Lanczos performs at least as well as explicit polynomial methods (for instance the kernel polynomial method \cite{kpm_review_06}); see \cite{chen_trogdon_ubaru_22} for a discussion and comparison.

While we are aware of no similarly rigorous analyses for the block Lanczos algorithm, we believe it is reasonable that similar results hold, albeit with possibly worse dependencies on certain parameters.
Our numerical experiments suggest the iterate $\vec{E}_1^\cT f[\Tl] \vec{E}_1$ computed by the block Lanczos algorithm still provides a good approximation to $\vec{V}^\cT f[\vec{H}] \vec{V}$, even when orthogonality of the Lanczos vectors is lost.
A rigorous analysis of the block Lanczos algorithm in floating point arithmetic is needed in order to make any definitive statements.

\section{Numerical experiments}

In this section we provide several numerical examples to demonstrate the accuracy and flexibility of our approach.
In all cases we consider isotropic XY spin systems; i.e. $J_{i,j}^{\textup{x}} = J_{i,j}^{\textup{y}}$ and $J_{i,j}^{\textup{z}} = 0$.

\subsection{Solvable system}

\begin{figure}
    \centering
    \includegraphics[width=.45\textwidth]{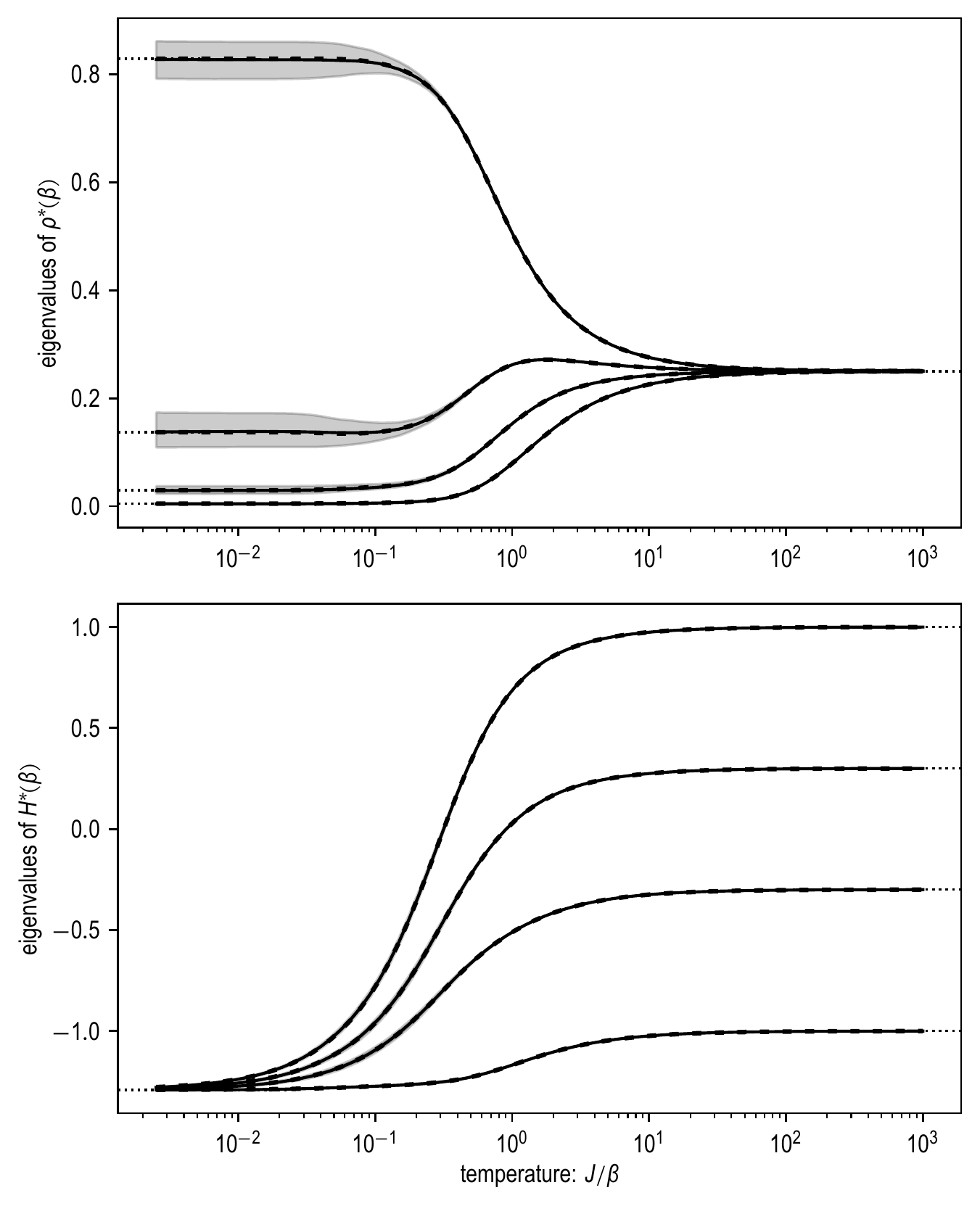}
    \caption{Eigenvalues of $\vec{H}^*(\beta)$ and $\rho^*(\beta)$ for a spin chain of length $18$ with system taken as the first two sites.
    \emph{Legend}:
    Algorithm median
    ({\protect\raisebox{0mm}{\protect\includegraphics[scale=.7]{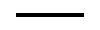}}}),
    Algorithm 10\%-90\% quantiles
    ({\protect\raisebox{0mm}{\protect\includegraphics[scale=.7]{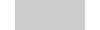}}}),
    Exact solution
    ({\protect\raisebox{0mm}{\protect\includegraphics[scale=.7]{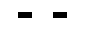}}}),
    Direct numerical computation of high and low temperature limits
    ({\protect\raisebox{0mm}{\protect\includegraphics[scale=.7]{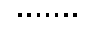}}}).
    }
    \label{fig:chain}
\end{figure}

We begin with the simple nearest neighbor spin chain with connection strength $J$.
We set $N = 18$, take the system to be the 1st and 2nd spin in the chain ($N_{\sys} = 2$), and put the magnetic field strength at $h=0.3J$.
We then run our algorithm using $k=30$ Lanczos iterations and $\nv = 100$ samples. 

\Cref{fig:chain}, shows the median, $10\%$ quantile, and $90\%$ quantile of 100 independent runs of our algorithm with the above parameters.
Because the spin chain is solvable analytically, we are able to compare our results against the exact eigenvalues of $\rho^*(\beta)$ and $\vec{H}^*(\beta)$ which were computed in \cite{campisi_zueco_talkner_10} and are summarized in \cref{sec:spin_chain_sol}.

Note that the quality of the approximation of the eigenvalues of $\rho^*(\beta)$ is accurate visually except at lower temperatures where there is higher variance (although the median outputs of our algorithm still agree very well with the true values).
The higher variance as low temperature expected based on our analysis above and could be decreased by increasing $\nv$.
However, at low temperature, the eigenvalues of $\rho^*(\beta)$ can be easily obtained by directly computing the ground state of the total system. 

\begin{figure*}[htb]
    \centering
    \includegraphics[width=\textwidth]{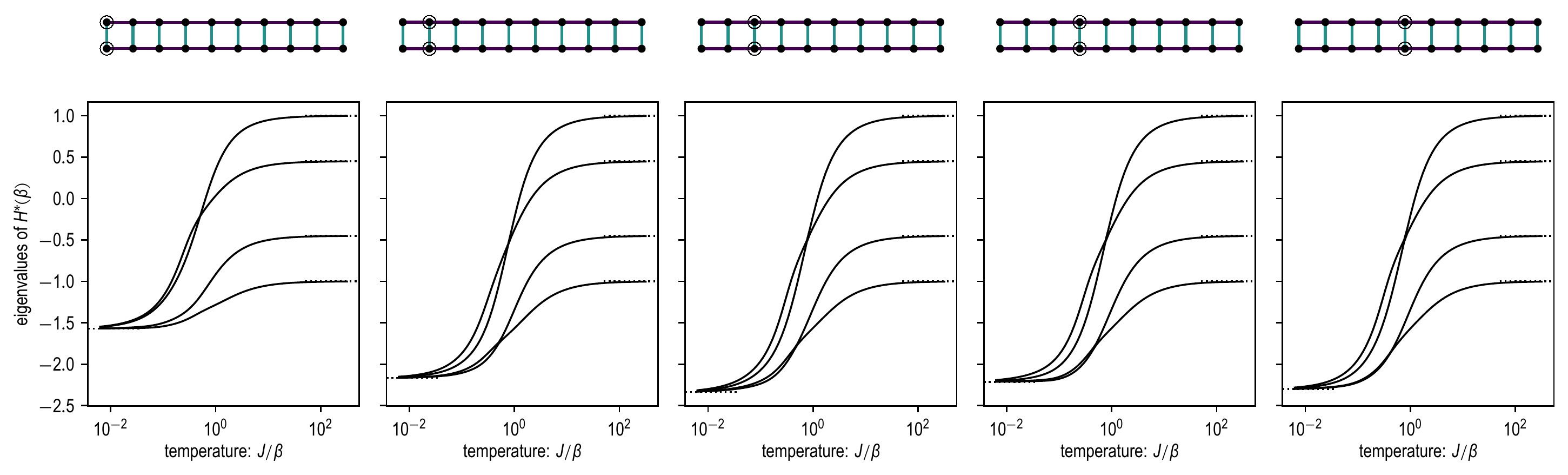}
    \caption{Eigenvalues of $\vec{H}^*(\beta)$ for a spin ladder with $10$ rungs with various choices of system (shown as circled sites in depiction of system configuration above images).
    \emph{Legend}:
    Algorithm
    ({\protect\raisebox{0mm}{\protect\includegraphics[scale=.7]{solid.pdf}}}),
    Direct numerical computation of high and low temperature limits
    ({\protect\raisebox{0mm}{\protect\includegraphics[scale=.7]{dot.pdf}}}).
    }
    \label{fig:ladder}
\end{figure*}

For all temperatures observed, we see a good agreement between our algorithm's approximation to the eigenvalues of $\vec{H}^*(\beta)$ and the exact eigenvalues of $\vec{H}^*(\beta)$.
As expected, at high temperature we observe that the spectrum of $\vec{H}^*(\beta)$ matches the spectrum of $\vec{H}_{\sys}$ while at low temperature the spectrum converges to a constant $E_{\tot} - E_{\bath}$.

\subsection{Varying choice of system spins}

We now consider a spin ladder.
The edge coupling strength is set to $J$, the run coupling strength is set to $-.45J$, and the magnetic field strength is set to $J$.
We set $N = 20$ and consider systems of size $N_{\sys}=2$ of spins connected by a rung.
There are 10 such systems, although only 5 are unique due to symmetry. 

In \cref{fig:ladder} we show the temperature dependence of the mean force Hamiltonian's eigenvalues for these 5 choices of system computed using $k=30$ and $\nv = 50$.
Because the bare system is the same in all cases, the high temperature behavior is the same. 
However, the low temperature limit as well as the qualitative behavior at intermediate temperature are different.

We observe the eigenvalues of the mean force Hamiltonian appear to ``cross'' implying  the occurrence of degenerate energy levels in the effective Hamiltonian at certain temperatures. 
More precisely, if the system starts from high enough temperature, we can guarantee all energy levels are not degenerate. 
Then when the temperature decreases to a certain point, degeneracy appears but it immediately disappears right after passing that temperature. 
Eventually, those energy levels converge to a single level in the zero temperature limit. We call this phenomenon ``temperature-induced degeneracy". 
This phenomenon is impossible for weak interaction systems ($H_{\syba} = \vec{\mathit{0}}$) since only strong interaction systems have a temperature-dependent effective Hamiltonian.
We suggest that there might be a certain unspecified type of symmetry induced by strong interactions at certain temperature so that the degeneracy appears due to that symmetry. 
We believe that specifying that strong interaction-induced symmetry and its connection with degeneracy is worthy of a further work.

\subsection{Long range interactions}

Next, we turn our attention to a spin chain with long range interactions in the presence of magnetic fields of varying strength. 
Specifically, for a spin chain with $N=16$ spins, we take $J_{i,j}^{\textup{x}} = J_{i,j}^{\textup{y}} = J |i-j|^{-\alpha}$, $\alpha\geq 0$. 
While $\alpha =\infty$ gives the solvable system studied above, to the best of our knowledge, this system is not exactly solvable for arbitrary $\alpha$.
Throughout, the system is taken as the first two spins and the bath as the latter 14.

\subsubsection{von Neumann entropy}

\begin{figure}[htb]
    \centering
    \includegraphics[width=.5\textwidth]{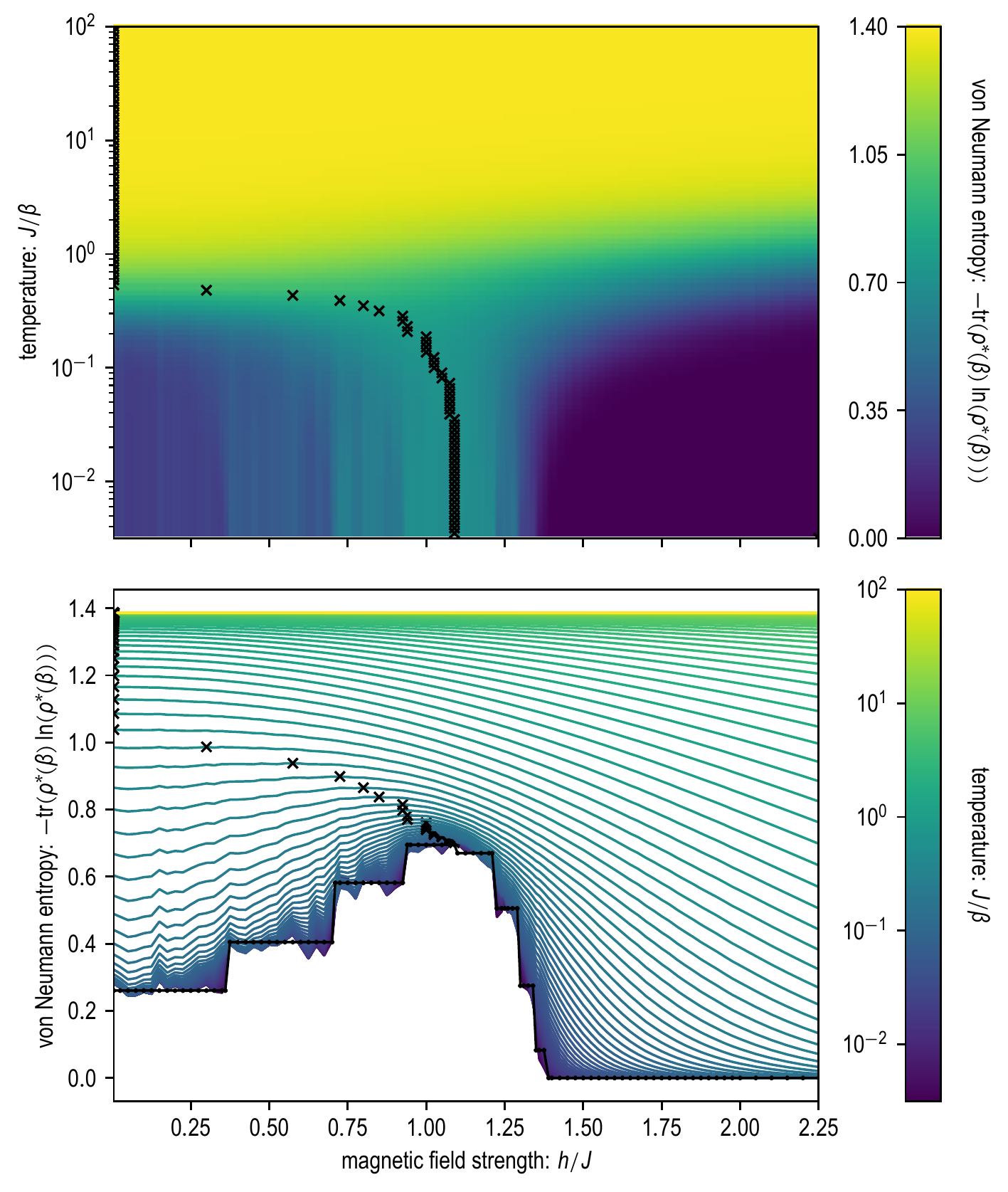}
    \caption{Relationship between von Neumann entropy, temperature, and magnetic field strength in spin chain with with long range power law interactions.
    Here, the system is taken as the first two spins and the bath as the remaining spins.
    Maximum value (at a fixed temperature) shows as black ``$\times$''.}
    \label{fig:vNEntropy}
\end{figure}

For this experiment, we set $\alpha = 1$ and vary $h$ from $0J$ to $2.2J$. 
We run our algorithm with $\nv = 400$ and $k=60$ and note that each value of $h$ requires an entirely independent run of our algorithm.

Phase plots showing the relationship between the system's von Neumann entropy $-\tr(\rho^*(\beta)\ln(\rho^*(\beta)))$, temperature, and magnetic field strength are given in \cref{fig:vNEntropy}.
The zero temperature limit is computed by using a black box eigensolver to find the total system ground state of the total system (under the assumption of a single ground state).
The relative smoothness between consecutive values of magnetic field strength provides some indication of the variance of the output produced by our algorithm.

At zero temperature, the von Neumann entropy indicates the presence of a magnetic field-induced quantum phase transition \cite{rost_perry_mercure_mackenzie_grigera_09, werlang_troppe_ribeiro_rigolin_10,  koffel_lewenstein_tagliacozzo_12, breunig_garst_klumper_rohrkamp_turnbull_lorenz_17} from an entangled state to a pure state  of the system of interest around $h \approx 1.4J$.
In the exactly solvable chain with $\alpha = \infty$, a similar quantum phase transition from entangled to pure state occurs at $h=2J$ \cite{campisi_zueco_talkner_10}.
This implies a dependence of the location of the quantum phase transition on the interaction range.
This dependence is studied for a closely related model in  \cite{koffel_lewenstein_tagliacozzo_12}.

Since zero temperature is not experimentally realizable, practical studies of quantum phase transitions require observations to be made at finite temperature \cite{sondhi_girvin_carini_97,werlang_troppe_ribeiro_rigolin_10}.
For $\alpha=1$, we find a region corresponding to low-entanglement extending beyond $T=0$ when $h$ is sufficiently large.
Despite some finite sample size noise in the output of our algorithm, it is clear that the staircase-like behavior observed at zero temperature extends to finite temperature.
These same phenomena is present in the $\alpha=\infty$ case, and a comparison of the $\alpha = 1$ phase plots in \cref{fig:vNEntropy} with the $\alpha=\infty$ phase plots in \cref{fig:vNEntropy_solvable} in the appendix shows a dependence on the interaction range.

Based on the above finding, we believe that our algorithm, which enables the temperature dependence of the von Neumann entropy (and similar quantities) to be studied at finite temperature, has the potential to inform the study of quantum phase transitions.

In addition to the magnetic field-induced quantum phase transition, we also observe a temperature-induced phase transition by observing the strength of the magnetic field at which the von Neumann entropy is maximal (at a fixed temperature). 
The value of these maxima are shown as black ``$\times$'' in \cref{fig:vNEntropy} and show the emergence of a critical phenomenon: the maximal von Neumann entropy at high temperature is obtained at zero magnetic field strength.
However, for temperatures below a critical temperature $T_{\textup{c}} \approx 0.5J/\beta$, the maximal von Neumann entropy occurs at nonzero magnetic field strength.

\subsubsection{Deviation in internal energy}
From \eqref{eqn:total_Hamiltonian_decomp}, it is not instantly clear how to split the energy due to $\vec{H}_{\syba}$ into the system of interest and the bath \cite{jarzynski_17}. Here, we adopt a difference of state functions (equilibrium averages of fluctuating observables) \emph{before and after} coupling to answer this question.
Before coupling, $\tr(H_{\sys} \rho_{\sys})$ is the equilibrium average of the internal energy of the system. 
On the other hand, after coupling, the relevant equilibrium average for the internal energy of the system is $\tr(H^*(\beta) \rho^*(\beta))$.
Therefore, the difference
\begin{equation}
\label{formula:deviation_of_Hamiltonian}
    \tr(H^*(\beta) \rho^*(\beta)) -  \tr(H_{\sys} \rho_{\sys}),
\end{equation}
can be interpreted as the ``deviation'' in the system's internal energy (state function) by coupling to the bath via $\vec{H}_{\syba}$. 

We note that the state functions of internal energy given in \cref{formula:deviation_of_Hamiltonian} are based on the approach by mean energy rather than the approach by partition function.
For strongly coupled systems, these two approaches lead to different thermodynamic results \cite{gelin_thoss_09, seifert_16, hsiang_hu_18, talkner_hanggi_20}, which are not in the scope of this work. 
On the other hand, the deviation defined by the difference in \cref{formula:deviation_of_Hamiltonian} is more relevant to the solvation energy generated in a transfer process of taking a solute from a vacuum to a solution \cite{arieh_13}. 

\begin{figure}
    \centering
    \includegraphics[width=.5\textwidth]{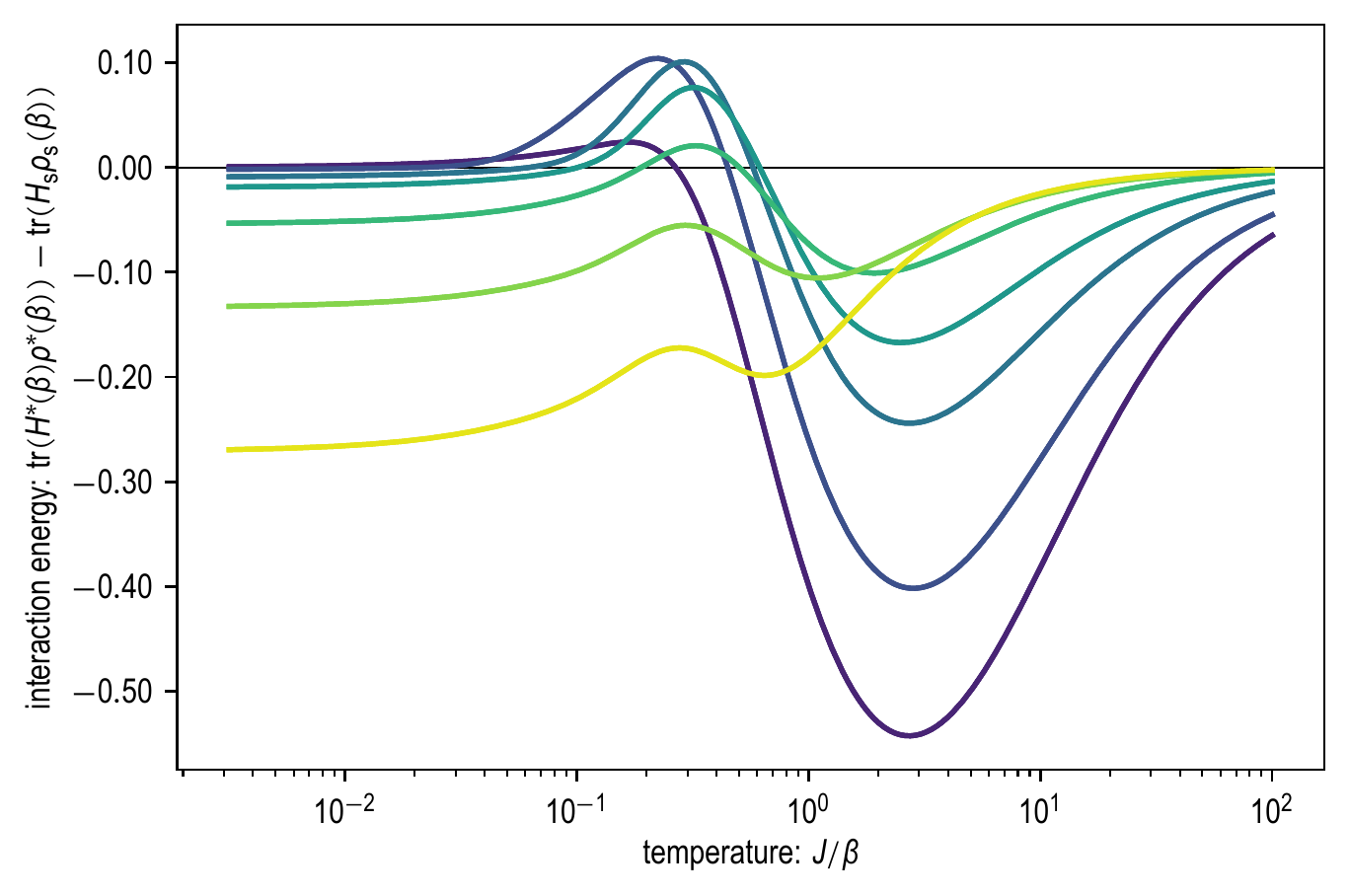}
    \caption{Difference of $\tr(H^*(\beta) \rho^*(\beta))$ and $\tr(H_{\sys} \rho_{\sys}(\beta))$ for a spin chain with varying interaction decay rates.
    In all cases, the system is taken as the first two spins and the bath as the remaining spins.
    \emph{Legend}:
    $\alpha = 0$ ({\protect\raisebox{0mm}{\protect\includegraphics[scale=.7]{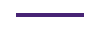}}}), 
    $\alpha = .1$ ({\protect\raisebox{0mm}{\protect\includegraphics[scale=.7]{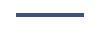}}}), 
    $\alpha = .3$ ({\protect\raisebox{0mm}{\protect\includegraphics[scale=.7]{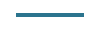}}}), 
    $\alpha = .5$ ({\protect\raisebox{0mm}{\protect\includegraphics[scale=.7]{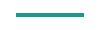}}}), 
    $\alpha = 1$ ({\protect\raisebox{0mm}{\protect\includegraphics[scale=.7]{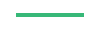}}}), 
    $\alpha = 2$ ({\protect\raisebox{0mm}{\protect\includegraphics[scale=.7]{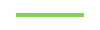}}}), and
    $\alpha = \infty$ ({\protect\raisebox{0mm}{\protect\includegraphics[scale=.7]{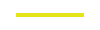}}}).
    }
    \label{fig:energy_alpha}
\end{figure}

To study this quantity we set $h=0$ and vary $\alpha$, the parameter controlling the interaction range. 
We run our algorithm with $\nv = 100$ and $k=60$. 
As seen in \cref{fig:energy_alpha}, the deviation is dependent on $\alpha$.
Specifically, we observe that at low temperatures, shorter range interactions (larger $\alpha$) correspond to a higher deviation, while at high temperatures, longer range interactions (lower $\alpha$) correspond to a higher deviation.
Moreover, we observe that the deviation of internal energy for longer-range interactions changes sign from positive to negative in certain temperature regions. 

Borrowing the idea from solvation thermodynamics \cite{arieh_13} and hybridization processes \cite{gong2020equilibrium}, our observation suggests that the coupling process for the system who has the longer-range interactions with its coupled bath can be either an endothermic process (positive regions) or an exothermic process (negative regions), and it is determined by the temperature! This further implies that the  system and its bath switches their interactions from being attractive to begin repelling or {\em vice versa} at certain critical temperatures. Our simulations also indicate the existence of a critical point  $\alpha_{\textup{c}} \in (1, 2)$ such that deviation is negative at all temperatures if $\alpha > \alpha_{\textup{c}}$, i.e., the coupling process for the system is always exothermic when the range of interactions is too short. 

\begin{figure}[htb]
    \centering
    \includegraphics[width=.5\textwidth]{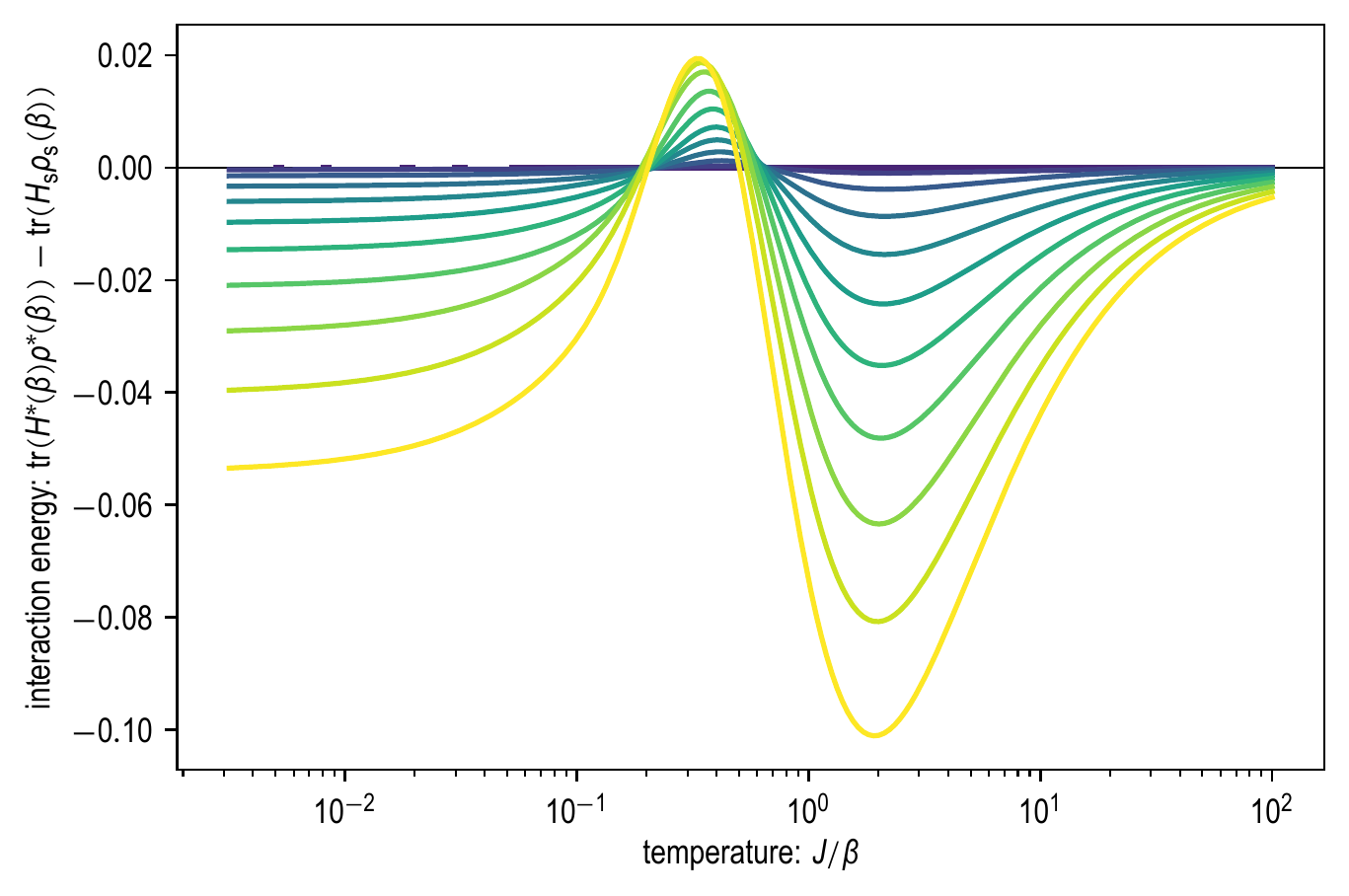}
    \caption{$\tr(H^*(\beta) \rho^*(\beta))$ and $\tr(H_{\sys} \rho_{\sys}(\beta))$ for a spin chain with long range power law interactions and varying system-bath coupling strengths.
    In all cases, the system is taken as the first two spins and the bath as the remaining spins.
    \emph{Legend}:
    $\epsilon = 0$ ({\protect\raisebox{0mm}{\protect\includegraphics[scale=.7]{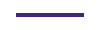}}})
    to $\epsilon = 1$ 
    ({\protect\raisebox{0mm}{\protect\includegraphics[scale=.7]{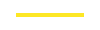}}})
    in increments of $0.1$.
    }
    \label{fig:energy_strong_weak}
\end{figure}

\subsection{Strong to weak coupling}

We now study the effect of the coupling strength on the deviation of the long range spin chain from the previous example when $\alpha = 1$ and $h=0$.
Specifically, we consider the Hamiltonian
\begin{equation}
    \vec{H}_{\tot} = \vec{\bar H}_{\sys} + \vec{\bar H}_{\bath} +  \epsilon \vec{H}_{\syba}
\end{equation}
where $\epsilon\geq0$ determines the coupling strength and $\vec{\bar H}_{\sys}$, $\vec{\bar H}_{\bath}$, and $\vec{H}_{\syba}$ are all as in the previous example.
We again run our algorithm with $\nv = 100$ and $k=60$.

In \cref{fig:energy_strong_weak} we observe that as the coupling strength decreases, the deviation also decreases. 
In particular, in the limit $\epsilon = 0$, the deviation is zero. 
This aligns with the limits described in \cref{sec:limits}.
While the deviation depends on the $\epsilon$, we observe a change in sign for all $\epsilon>0$). 
This suggests the existence of temperature-induced sign changes in the energy deviation depends on the range, rather than the strength, of interactions.

\section{Conclusion}

We have introduced a numerical algorithm, based on the concept of partial typicality, for computing the mean force Hamiltonian and average reduced system density matrix of strongly coupled spin systems. 
Numerical experiments on a solvable system indicate that our algorithm can produce highly accurate results.
This is complemented by a theoretical analysis of the behavior of the algorithm. 
Further experiments on a range of systems which are not exactly solvable demonstrate the flexibility and power of the algorithm.
We hope that this algorithm will enable further study of the thermodynamic properties of open quantum systems. 

There are a number of concrete directions for future work. 
First, it would be interesting to extend methods which balance low-rank approximation and stochastic trace estimation \cite{meyer_musco_musco_woodruff_21,persson_cortinovis_kressner_22,chen_hallman_22} to the task of computing partial traces of matrix functions.
In particular, the recent approach of \cite{chen_hallman_22} demonstrates the effectiveness of hybrid approaches for matrix function trace estimation problems in quantum physics.
Generalizing the approach to computing partial traces would enable higher quality approximations at low temperature.
It would also be interesting to study how variants of the finite temperature Lanczos method, such as the microcanonical Lanczos method (MCLM)\cite{long_prelovsek_elshawish_karadamoglou_zotos_03} and low temperature Lanczos method (LTLM)\cite{aichhorn_daghofer_evertz_vondelinden_03}, can be extended to the setting of this paper.
Finally, we believe it should be relatively straightforward to use the same techniques used in this paper, typicality for partial traces and block Krylov subspace methods, to compute dynamical quantities.

\section*{Acknowledgments}

The authors thank Michele Campisi, Hong Qian, Peter Talkner, Lowell Thompson, Yao Wang, Yijing Yan, and Ying-Jen Yang for feedback and suggestions.

This material is based on work supported by the National Science Foundation under Grant No. DGE-1762114. Any opinions, findings, and conclusions or recommendations expressed in this material are those of the authors and do not necessarily reflect the views of the National Science Foundation.

\section*{Author  Declarations}

\subsection*{Conflict of interest}
\noindent
The authors have no conflicts to disclose.

\subsection*{Data availability}
\noindent
The data that support the findings of this study are available within the article.

\appendix
\section{Derivation of high and low temperature limits}
\label{sec:Hmf_limit}

\subsection{High temperature limit}
Expanding at $\beta = 0$ we find
\begin{align}
    \hspace{2em}&\hspace{-2em}
    \ln \left[  \tr_{\bath}(\exp[ -\beta \vec{H}_{\tot}]) \right]
    \\&= \ln \left[  \tr_{\bath}(\Id_{\sys}\otimes \Id_{\bath} -\beta \vec{H} + \mathcal{O}(\beta^2)) \right] 
    \\&= \ln \left[  \tr(\Id_{\bath}) \Id_{\sys} -\beta \tr_{\bath}( \vec{H} ) + \mathcal{O}(\beta^2) \right] 
    \\&= \ln \left[  \Id_{\sys} - \beta \frac{\tr_{\bath}(\vec{H}_{\tot})}{\tr(\Id_{\bath})} + \mathcal{O}(\beta^2) \right] + \ln[ \tr(\Id_{\bath}) \Id_{\sys} ]
    \\&= - \beta \frac{\tr_{\bath}(\vec{H}_{\tot})}{\tr(\Id_{\bath})} + \mathcal{O}(\beta^2) + \ln( \tr(\Id_{\bath}) ) \Id_{\sys} .
\end{align}
Here we have used the property that $\ln[\vec{A}\vec{B}] = \ln[\vec{A}] + \ln[\vec{B}]$ if $\vec{A}$ and $\vec{B}$ commute.
Similarly, we can expand
\begin{align}
    \ln(\tr(\exp[-\beta \vec{H}_{\bath}]))
    &= \ln(\tr(\vec{I}_{\bath} - \beta \vec{H}_{\bath} + \mathcal{O}(\beta^2)))
    \\&= -\beta \frac{\tr(\vec{H}_{\bath})}{\tr(\vec{I}_{\bath})} + \mathcal{O}(\beta^2) + \ln(\tr(\vec{I}_{\bath})).
\end{align}
Thus, combining these expressions,
\begin{align}
    H^*(\beta) \hspace{-3em}&\hspace{3em}= -\frac{1}{\beta}\ln \left[ \frac{ \tr_{\bath}(\exp[ -\beta \vec{H}_{\tot}])}{\tr( \exp[-\beta \vec{H}_{\bath}] )} \right]  
    \\&=  -\frac{1}{\beta} \ln \left[  \tr_{\bath}(\exp[ -\beta \vec{H}_{\tot}]) \right] \nonumber
    \\&\hspace{3em}+ \frac{1}{\beta}  \ln \left[ \tr( \exp[-\beta \vec{H}_{\bath}] ) \right] \Id_{\sys}
    \\&= \frac{\tr_{\bath}(\vec{H}_{\tot})}{\tr(\Id_{\bath})} - \frac{\tr(\vec{H}_{\bath})}{\tr(\Id_{\bath})} \Id_{\sys} + \mathcal{O}(\beta) 
    \\&= \vec{H}_{\sys} + \frac{\tr_{\bath}(\vec{H}_{\syba})}{\tr(\Id_{\bath})} + \mathcal{O}(\beta),
\end{align}
where, in the final equality, we have used that 
\begin{align}
    \tr_{\bath}(\vec{H}_{\tot})
    &=  \tr_{\bath}( \vec{H}_{\sys} \otimes \Id_{\bath} + \Id_{\sys}\otimes \vec{H}_{\bath} + \vec{H}_{\syba} )
    \\&= \tr(\Id_{\bath}) \vec{H}_{\sys} + \tr(\vec{H}_{\bath}) \Id_{\sys} + \tr_{\bath}(\vec{H}_{\syba}).
\end{align}
Now, recall that $\vec{H}_{\syba}$ accounts for interactions between the system and bath and is therefore a linear combination of of terms of the form
\begin{align}
&( \Id_{i} \otimes \vec{\sigma}^{\textup{x}/\textup{y}/\textup{z}} \otimes \Id_{i'})\otimes \Id_{\bath})(\Id_{\sys}\otimes(\Id_{j}\otimes \vec{\sigma}^{\textup{x}/\textup{y}/\textup{z}}\otimes \Id_{j'})
\\&\hspace{4em}
= ( \Id_{i} \otimes \vec{\sigma}^{\textup{x}/\textup{y}/\textup{z}} \otimes \Id_{i'}) \otimes (\Id_{j}\otimes \vec{\sigma}^{\textup{x}/\textup{y}/\textup{z}}\otimes \Id_{j'}),
\end{align}
where $i+i' = (2s+1)^{N_\sys-1}$ and $j+j'= (2s+1)^{N_\bath-1}$.
Applying basic properties of the partial trace  we see that
\begin{align}
    \hspace{3em}&\hspace{-3em}\tr_{\bath}\big(( \Id_{i} \otimes \vec{\sigma}^{\textup{x}/\textup{y}/\textup{z}} \otimes \Id_{i'}) \otimes (\Id_{j}\otimes \vec{\sigma}^{\textup{x}/\textup{y}/\textup{z}}\otimes \Id_{j'})\big) 
    \\&= \tr_{\bath}\big( \Id_{j}\otimes \vec{\sigma}^{\textup{x}/\textup{y}/\textup{z}}\otimes \Id_{j'}\big) ( \Id_{i} \otimes \vec{\sigma}^{\textup{x}/\textup{y}/\textup{z}} \otimes \Id_{i'}) 
    \\&= \tr_{\bath}(\Id_{j}) \tr_{\bath}(\vec{\sigma}^{\textup{x}/\textup{y}/\textup{z}}) \tr_{\bath}(\Id_{j'}) ( \Id_{i} \otimes \vec{\sigma}^{\textup{x}/\textup{y}/\textup{z}} \otimes \Id_{i'}) 
    \\&= \vec{\mathit{0}}_{\sys}.
\end{align}
Here we have used the fact that $\tr(\vec{\sigma}^{\textup{x}/\textup{y}/\textup{z}}) = 0$ for any spin number $s$.Numerical computation of the reduced density matrix
The trace is linear, so we in fact have that $\tr_{\bath}(\vec{H}_{\syba}) = \vec{\mathit{0}}_{\sys}$.

We have therefore established that
\begin{equation}
    \lim_{\beta\to 0} \vec{H}^*(\beta) \to \vec{H}_{\sys}.
\end{equation}

\subsection{Low temperature limit}

Write the eigenvalue decomposition of $\vec{H}_{\tot}$ as $\vec{H}_{\tot} = \sum_{i=1}^{|\mathcal{H}_\tot|} E_i | \vec{\psi}_i \rangle \langle \vec{\psi}_i |$ for orthogonal eigenvectors $| \vec{\psi}_i \rangle$. 
Then, 
\begin{align}
    \rho(\vec{H}_{\tot})
    = \exp[-\beta \vec{H}_{\tot}]
    = \frac{\sum_{i=1}^{|\mathcal{H}_\tot|} \exp(-\beta E_i) | \vec{\psi}_i \rangle \langle \vec{\psi}_i |}{\sum_{i=1}^{|\mathcal{H}_\tot|} \exp(-\beta E_i)}.
\end{align}
When $\beta \to\infty$ only the terms corresponding to the ground state (smallest absolute eigenvalue) remain. 

Note that
\begin{align}
    \vec{H}^*(\beta) 
    &= -\frac{1}{\beta} \ln[Z^*(\beta)\vec{\rho}^*(\beta)]
    \\&= -\frac{1}{\beta} \ln(Z^*(\beta)) \Id_{\sys} - \beta^{-1} \ln[\vec{\rho}^*(\beta)].
\end{align}
Since $\vec{\rho}^*(\beta)$ is convergent to a fixed density matrix, $\beta^{-1} \ln[\vec{\rho}^*(\beta)]\to \vec{\mathit{0}}_{\sys}$ as $\beta\to\infty$.
By definition, 
\begin{align}
    -\frac{1}{\beta} \ln(Z^*(\beta))
    &= -\frac{1}{\beta} \ln(Z_{\tot}(\beta) / Z_{\bath}(\beta)) 
    \\&= -\frac{1}{\beta} \ln(Z_{\tot}(\beta)) + \frac{1}{\beta} \ln( Z_{\bath}(\beta)).
\end{align}
In the low temperature limit, the partition functions for the bare system and bare bath are dominated by their respective ground state energies so that
\begin{align}
-\frac{1}{\beta} \ln(Z_{\tot/\bath}(\beta))
&= -\frac{1}{\beta} \ln(\tr(\exp(-\beta \vec{H}_{\tot/\bath}))) 
\\&= \lambda_{\textup{max}}( \vec{H}_{\tot/\bath} )  + \mathcal{O}(\beta^{-1}).
\end{align}
We therefore have that
\begin{equation}
    \lim_{\beta \to \infty} \vec{H}^*(\beta) = (E_{\tot} - E_{\bath}) \Id_{\sys}
\end{equation}
where $E_{\tot}$ and $E_{\bath}$ are the ground state energies of $\vec{H}_{\tot}$ and $\vec{H}_{\bath}$ respectively.

\section{A consistent partial trace estimator}

Consider a random pure product state $| \vec{v}_{\sys} \rangle \otimes | \vec{v}_{\bath} \rangle$ where $\HA[|\vec{v}_{\sys/\bath}\rangle\langle \vec{v}_{\sys/\bath}|] = \Id_{\sys/\bath}$.
We have that
\begin{align}
    &\langle \vec{v}_{\sys} | ( \Id_{\sys} \otimes \langle \vec{v}_{\bath} | )  \vec{A} (\Id_{\sys} \otimes | \vec{v}_{\bath} \rangle ) | \vec{v}_{\sys} \rangle 
    \nonumber
    \\&\hspace{4em}=  (\langle \vec{v}_{\sys} | \otimes \langle \vec{v}_{\bath} |) \vec{A} (| \vec{v}_{\sys} \rangle \otimes | \vec{v}_{\bath} \rangle)
    \\&\hspace{4em}= (\Id_{\bath} \otimes \langle \vec{v}_{\sys} |) \langle \vec{v}_{\bath} | )  \vec{A} | \vec{v}_{\bath} \rangle  | (\Id_{\bath} \otimes \langle \vec{v}_{\sys} |) 
\end{align}
Thus, estimators of this form have the desirable property that
\begin{align}
    \HA\big[ (\langle \vec{v}_{\sys} | \otimes \langle \vec{v}_{\bath} |) \vec{A} (| \vec{v}_{\sys} \rangle \otimes | \vec{v}_{\bath} \rangle) \:\big|\: \vec{v}_{\sys} \big] 
    &= \langle \vec{v}_{\sys} | \tr_{\bath} ( \vec{A} ) | \vec{v}_{\sys} \rangle
    \\
    \HA\big[ (\langle \vec{v}_{\bath} | \otimes \langle \vec{v}_{\sys} |) \vec{A} (| \vec{v}_{\bath} \rangle \otimes | \vec{v}_{\sys} \rangle) \:\big|\: \vec{v}_{\bath} \big] 
    &= \langle \vec{v}_{\bath} | \tr_{\sys} ( \vec{A} ) | \vec{v}_{\bath} \rangle.
\end{align}
That is, we obtain approximations of the trace and partial trace which are consistent in the sense that our estimate of the trace of the partial traces are equal. 

Expressions of the form $| \vec{v}_{\sys} \rangle \otimes | \vec{v}_{\bath} \rangle$ are called rank-one vectors and have been studied for the task of trace estimation with the goal of reducing the amount of randomness required \cite{bujanovic_kressner_21}.
It is conceivable that there are situations in which using a pure product state $|\vec{v} \rangle = |\vec{v}_1 \rangle\otimes \cdots \otimes |\vec{v}_N \rangle$ would be desirable.
While this produces unbiased estimators for arbitrary partial traces, the number of such estimators which must be averaged to reach a fixed error is exponential in $N$ \cite{vershynin_20,bamberger_krahmer_ward_21}.

\section{Solution to the spin chain}
\label{sec:spin_chain_sol}

We summarize the relevant quantities from \cite{campisi_zueco_talkner_10}.
For $k=1, \ldots, N$, define
\begin{align}
    \lambda_k^{(N)} 
    &= h - 2 J \cos \left( \frac{k\pi}{N+1} \right)
    \\
    \mathcal{N}_k^{(N)} 
    &= (1 + \exp(\beta \lambda_k^{(N)} )^{-1}.
\end{align}
Next, define 
\begin{align}
    \langle \sigma_1^\textup{x}\sigma_2^\textup{x} \rangle
    &= \frac{-4}{N+1} \sum_{k=1}^{N} \sin \left( \frac{k\pi}{N+1} \right) \sin \left( \frac{2k\pi}{N+1} \right) \mathcal{N}_k^{(N)}
    \\
    \langle \sigma_j^\textup{z} \rangle
    &= \frac{-4}{N+1} \sum_{k=1}^{N} \sin \left( \frac{jk\pi}{N+1} \right)^2 \mathcal{N}_k^{(N)}
    \\
    \langle \sigma_1^\textup{z}\sigma_2^\textup{z} \rangle
    &= \langle \sigma_1^\textup{z} \rangle \langle \sigma_2^\textup{z} \rangle - \langle \sigma_1^\textup{x}\sigma_2^\textup{x} \rangle^2
    \\
    \delta &= \sqrt{4 \langle \sigma_1^\textup{x}\sigma_2^\textup{x} \rangle^2 + (\sigma_1^\textup{z} \rangle - \langle \sigma_2^\textup{z} \rangle)^2}.
\end{align}
For $N_{\sys} = 2$, the eigenvalues of $\rho^*(\beta)$ are given by
\begin{align}
    p_1 &= (1 + \langle \sigma_1^\textup{z} \rangle+ \langle \sigma_2^\textup{z} \rangle+ \langle \sigma_1^\textup{z}\sigma_2^\textup{z} \rangle )/4
    \\
    p_2 &= (1-\delta - \langle \sigma_1^\textup{z}\sigma_2^\textup{z} \rangle) /4
    \\
    p_3 &= (1+\delta - \langle \sigma_1^\textup{z}\sigma_2^\textup{z} \rangle) / 4
    \\
    p_4 &= (1 - \langle \sigma_1^\textup{z} \rangle - \langle \sigma_2^\textup{z} \rangle+ \langle \sigma_1^\textup{z}\sigma_2^\textup{z} \rangle )/4.
\end{align}
Moreover, the partition function of a length $N$ chain is given by 
\begin{equation}
    Z_N = \exp(\beta N h / 2) \prod_{k=1}^{N} (1 + \exp(-\beta \lambda_k^{(N)} ).
\end{equation}
We have $Z^*(\beta) = Z_{N} / Z_{N_{\bath}}$, and the eigenvalues of $\vec{H}^*(\beta)$ are given by
\begin{equation}
    h_j =  -\frac{1}{\beta} \ln( Z^*(\beta) p_j )
    ,\qquad j=1,2,3,4.
\end{equation}

\section{Phase plot for solvable spin chain}

\begin{figure}[htb]
    \centering
    \includegraphics[width=.5\textwidth]{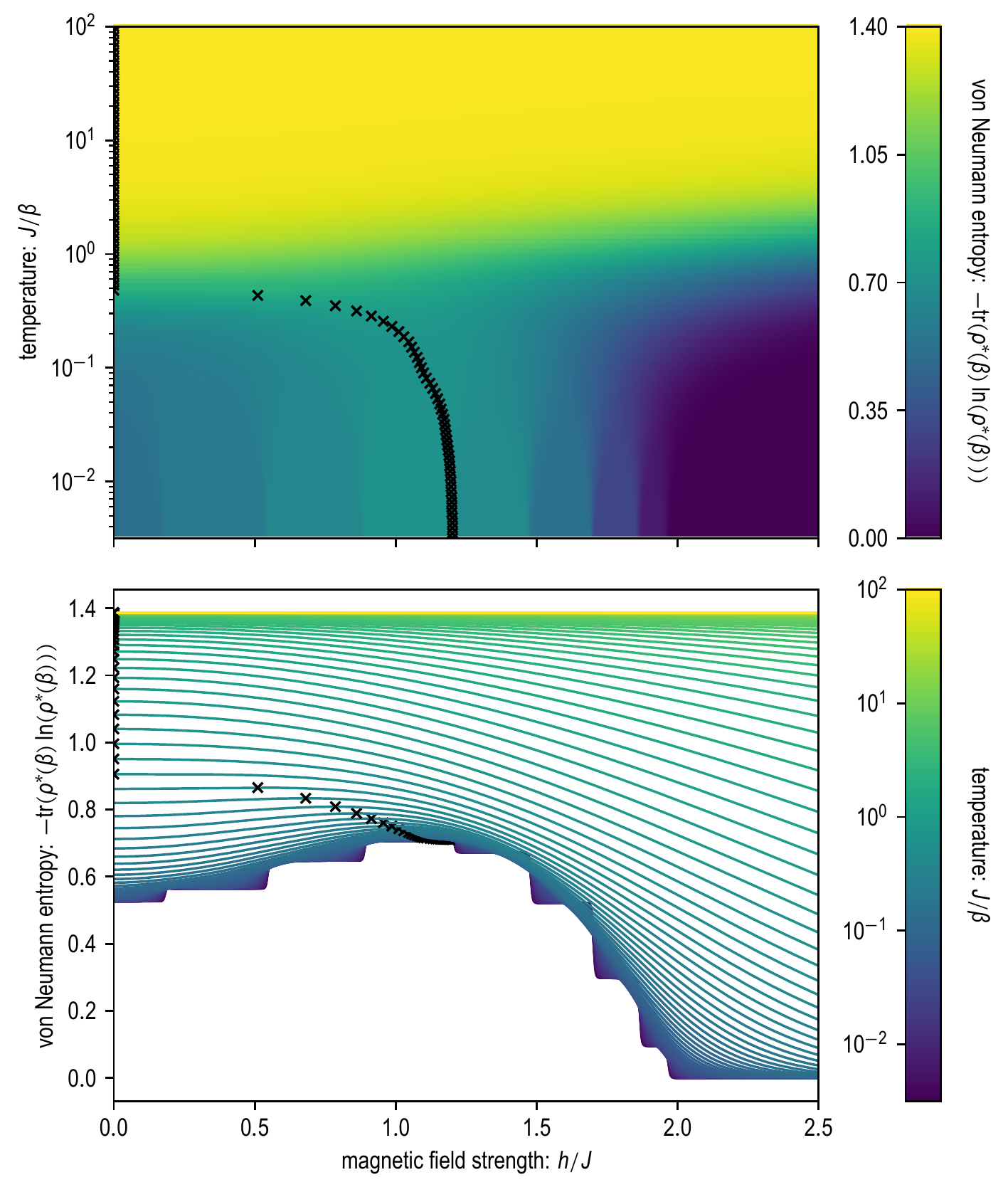}
    \caption{Relationship between von Neumann entropy, temperature, and magnetic field strength in spin chain with short range nearest-neighbor interactions.
    Here, the system is taken as the first two spins and the bath as the remaining spins.}
    \label{fig:vNEntropy_solvable}
\end{figure}

Phase plots showing the relationship between the  von Neumann entropy $-\tr(\rho^*(\beta)\ln(\rho^*(\beta)))$, temperature, and magnetic field strength for the exactly solvable spin chain with $N=16$ and the fist two spins taken as the system are given in \cref{fig:vNEntropy_solvable}.

\bibliography{refs}

\begin{thebibliography}{64}%
\makeatletter
\providecommand \@ifxundefined [1]{%
 \@ifx{#1\undefined}
}%
\providecommand \@ifnum [1]{%
 \ifnum #1\expandafter \@firstoftwo
 \else \expandafter \@secondoftwo
 \fi
}%
\providecommand \@ifx [1]{%
 \ifx #1\expandafter \@firstoftwo
 \else \expandafter \@secondoftwo
 \fi
}%
\providecommand \natexlab [1]{#1}%
\providecommand \enquote  [1]{``#1''}%
\providecommand \bibnamefont  [1]{#1}%
\providecommand \bibfnamefont [1]{#1}%
\providecommand \citenamefont [1]{#1}%
\providecommand \href@noop [0]{\@secondoftwo}%
\providecommand \href [0]{\begingroup \@sanitize@url \@href}%
\providecommand \@href[1]{\@@startlink{#1}\@@href}%
\providecommand \@@href[1]{\endgroup#1\@@endlink}%
\providecommand \@sanitize@url [0]{\catcode `\\12\catcode `\$12\catcode
  `\&12\catcode `\#12\catcode `\^12\catcode `\_12\catcode `\%12\relax}%
\providecommand \@@startlink[1]{}%
\providecommand \@@endlink[0]{}%
\providecommand \url  [0]{\begingroup\@sanitize@url \@url }%
\providecommand \@url [1]{\endgroup\@href {#1}{\urlprefix }}%
\providecommand \urlprefix  [0]{URL }%
\providecommand \Eprint [0]{\href }%
\providecommand \doibase [0]{http://dx.doi.org/}%
\providecommand \selectlanguage [0]{\@gobble}%
\providecommand \bibinfo  [0]{\@secondoftwo}%
\providecommand \bibfield  [0]{\@secondoftwo}%
\providecommand \translation [1]{[#1]}%
\providecommand \BibitemOpen [0]{}%
\providecommand \bibitemStop [0]{}%
\providecommand \bibitemNoStop [0]{.\EOS\space}%
\providecommand \EOS [0]{\spacefactor3000\relax}%
\providecommand \BibitemShut  [1]{\csname bibitem#1\endcsname}%
\let\auto@bib@innerbib\@empty
\bibitem [{\citenamefont {Gemmer}, \citenamefont {Michel},\ and\ \citenamefont
  {Mahler}(2009)}]{gemmer_michel_mahler_09}%
  \BibitemOpen
  \bibfield  {author} {\bibinfo {author} {\bibfnamefont {J.}~\bibnamefont
  {Gemmer}}, \bibinfo {author} {\bibfnamefont {M.}~\bibnamefont {Michel}}, \
  and\ \bibinfo {author} {\bibfnamefont {G.}~\bibnamefont {Mahler}},\
  }\href@noop {} {\emph {\bibinfo {title} {Quantum thermodynamics: emergence of
  thermodynamic behavior within composite quantum systems}}},\ \bibinfo
  {edition} {2nd}\ ed.,\ Lecture notes in physics\ (\bibinfo  {publisher}
  {Springer},\ \bibinfo {year} {2009})\BibitemShut {NoStop}%
\bibitem [{\citenamefont {Vinjanampathy}\ and\ \citenamefont
  {Anders}(2016)}]{vinjanampathy_anders_16}%
  \BibitemOpen
  \bibfield  {author} {\bibinfo {author} {\bibfnamefont {S.}~\bibnamefont
  {Vinjanampathy}}\ and\ \bibinfo {author} {\bibfnamefont {J.}~\bibnamefont
  {Anders}},\ }\bibfield  {title} {\enquote {\bibinfo {title} {Quantum
  thermodynamics},}\ }\href {\doibase 10.1080/00107514.2016.1201896} {\bibfield
   {journal} {\bibinfo  {journal} {Contemporary Physics}\ }\textbf {\bibinfo
  {volume} {57}},\ \bibinfo {pages} {545--579} (\bibinfo {year}
  {2016})}\BibitemShut {NoStop}%
\bibitem [{\citenamefont {Alicki}\ and\ \citenamefont
  {Kosloff}(2018)}]{alicki_kosloff_18}%
  \BibitemOpen
  \bibfield  {author} {\bibinfo {author} {\bibfnamefont {R.}~\bibnamefont
  {Alicki}}\ and\ \bibinfo {author} {\bibfnamefont {R.}~\bibnamefont
  {Kosloff}},\ }\bibfield  {title} {\enquote {\bibinfo {title} {Introduction to
  quantum thermodynamics: History and prospects},}\ }in\ \href {\doibase
  10.1007/978-3-319-99046-0_1} {\emph {\bibinfo {booktitle} {Fundamental
  Theories of Physics}}}\ (\bibinfo  {publisher} {Springer International
  Publishing},\ \bibinfo {year} {2018})\ pp.\ \bibinfo {pages}
  {1--33}\BibitemShut {NoStop}%
\bibitem [{\citenamefont {Ingold}, \citenamefont {H\"{a}nggi},\ and\
  \citenamefont {Talkner}(2009)}]{ingold_hanggi_talkner_09}%
  \BibitemOpen
  \bibfield  {author} {\bibinfo {author} {\bibfnamefont {G.-L.}\ \bibnamefont
  {Ingold}}, \bibinfo {author} {\bibfnamefont {P.}~\bibnamefont {H\"{a}nggi}},
  \ and\ \bibinfo {author} {\bibfnamefont {P.}~\bibnamefont {Talkner}},\
  }\bibfield  {title} {\enquote {\bibinfo {title} {Specific heat anomalies of
  open quantum systems},}\ }\href {\doibase 10.1103/physreve.79.061105}
  {\bibfield  {journal} {\bibinfo  {journal} {Physical Review E}\ }\textbf
  {\bibinfo {volume} {79}} (\bibinfo {year} {2009}),\
  10.1103/physreve.79.061105}\BibitemShut {NoStop}%
\bibitem [{\citenamefont {Campisi}, \citenamefont {Zueco},\ and\ \citenamefont
  {Talkner}(2010)}]{campisi_zueco_talkner_10}%
  \BibitemOpen
  \bibfield  {author} {\bibinfo {author} {\bibfnamefont {M.}~\bibnamefont
  {Campisi}}, \bibinfo {author} {\bibfnamefont {D.}~\bibnamefont {Zueco}}, \
  and\ \bibinfo {author} {\bibfnamefont {P.}~\bibnamefont {Talkner}},\
  }\bibfield  {title} {\enquote {\bibinfo {title} {Thermodynamic anomalies in
  open quantum systems: Strong coupling effects in the isotropic {XY} model},}\
  }\href {\doibase 10.1016/j.chemphys.2010.04.026} {\bibfield  {journal}
  {\bibinfo  {journal} {Chemical Physics}\ }\textbf {\bibinfo {volume} {375}},\
  \bibinfo {pages} {187--194} (\bibinfo {year} {2010})}\BibitemShut {NoStop}%
\bibitem [{\citenamefont {Talkner}\ and\ \citenamefont
  {H\"{a}nggi}(2020)}]{talkner_hanggi_20}%
  \BibitemOpen
  \bibfield  {author} {\bibinfo {author} {\bibfnamefont {P.}~\bibnamefont
  {Talkner}}\ and\ \bibinfo {author} {\bibfnamefont {P.}~\bibnamefont
  {H\"{a}nggi}},\ }\bibfield  {title} {\enquote {\bibinfo {title} {Colloquium :
  Statistical mechanics and thermodynamics at strong coupling: Quantum and
  classical},}\ }\href {\doibase 10.1103/revmodphys.92.041002} {\bibfield
  {journal} {\bibinfo  {journal} {Reviews of Modern Physics}\ }\textbf
  {\bibinfo {volume} {92}} (\bibinfo {year} {2020}),\
  10.1103/revmodphys.92.041002}\BibitemShut {NoStop}%
\bibitem [{\citenamefont {Makri}\ and\ \citenamefont
  {Makarov}(1995{\natexlab{a}})}]{makri_makarov_95}%
  \BibitemOpen
  \bibfield  {author} {\bibinfo {author} {\bibfnamefont {N.}~\bibnamefont
  {Makri}}\ and\ \bibinfo {author} {\bibfnamefont {D.~E.}\ \bibnamefont
  {Makarov}},\ }\bibfield  {title} {\enquote {\bibinfo {title} {Tensor
  propagator for iterative quantum time evolution of reduced density matrices.
  i. theory},}\ }\href {\doibase 10.1063/1.469508} {\bibfield  {journal}
  {\bibinfo  {journal} {The Journal of Chemical Physics}\ }\textbf {\bibinfo
  {volume} {102}},\ \bibinfo {pages} {4600--4610} (\bibinfo {year}
  {1995}{\natexlab{a}})}\BibitemShut {NoStop}%
\bibitem [{\citenamefont {Makri}\ and\ \citenamefont
  {Makarov}(1995{\natexlab{b}})}]{makri_makarov_95b}%
  \BibitemOpen
  \bibfield  {author} {\bibinfo {author} {\bibfnamefont {N.}~\bibnamefont
  {Makri}}\ and\ \bibinfo {author} {\bibfnamefont {D.~E.}\ \bibnamefont
  {Makarov}},\ }\bibfield  {title} {\enquote {\bibinfo {title} {Tensor
  propagator for iterative quantum time evolution of reduced density matrices.
  {II}. numerical methodology},}\ }\href {\doibase 10.1063/1.469509} {\bibfield
   {journal} {\bibinfo  {journal} {The Journal of Chemical Physics}\ }\textbf
  {\bibinfo {volume} {102}},\ \bibinfo {pages} {4611--4618} (\bibinfo {year}
  {1995}{\natexlab{b}})}\BibitemShut {NoStop}%
\bibitem [{\citenamefont {Gelzinis}\ and\ \citenamefont
  {Valkunas}(2020)}]{gelzinis_valkunas_20}%
  \BibitemOpen
  \bibfield  {author} {\bibinfo {author} {\bibfnamefont {A.}~\bibnamefont
  {Gelzinis}}\ and\ \bibinfo {author} {\bibfnamefont {L.}~\bibnamefont
  {Valkunas}},\ }\bibfield  {title} {\enquote {\bibinfo {title} {Analytical
  derivation of equilibrium state for open quantum system},}\ }\href {\doibase
  10.1063/1.5141519} {\bibfield  {journal} {\bibinfo  {journal} {The Journal of
  Chemical Physics}\ }\textbf {\bibinfo {volume} {152}},\ \bibinfo {pages}
  {051103} (\bibinfo {year} {2020})}\BibitemShut {NoStop}%
\bibitem [{\citenamefont {Trushechkin}\ \emph {et~al.}(2022)\citenamefont
  {Trushechkin}, \citenamefont {Merkli}, \citenamefont {Cresser},\ and\
  \citenamefont {Anders}}]{trushechkin_merkli_cresser_anders_22}%
  \BibitemOpen
  \bibfield  {author} {\bibinfo {author} {\bibfnamefont {A.~S.}\ \bibnamefont
  {Trushechkin}}, \bibinfo {author} {\bibfnamefont {M.}~\bibnamefont {Merkli}},
  \bibinfo {author} {\bibfnamefont {J.~D.}\ \bibnamefont {Cresser}}, \ and\
  \bibinfo {author} {\bibfnamefont {J.}~\bibnamefont {Anders}},\ }\bibfield
  {title} {\enquote {\bibinfo {title} {Open quantum system dynamics and the
  mean force gibbs state},}\ }\href {\doibase 10.1116/5.0073853} {\bibfield
  {journal} {\bibinfo  {journal} {{AVS} Quantum Science}\ }\textbf {\bibinfo
  {volume} {4}},\ \bibinfo {pages} {012301} (\bibinfo {year}
  {2022})}\BibitemShut {NoStop}%
\bibitem [{\citenamefont {Chiu}, \citenamefont {Strathearn},\ and\
  \citenamefont {Keeling}(2022)}]{chiu_strathearn_keeling_22}%
  \BibitemOpen
  \bibfield  {author} {\bibinfo {author} {\bibfnamefont {Y.-F.}\ \bibnamefont
  {Chiu}}, \bibinfo {author} {\bibfnamefont {A.}~\bibnamefont {Strathearn}}, \
  and\ \bibinfo {author} {\bibfnamefont {J.}~\bibnamefont {Keeling}},\
  }\bibfield  {title} {\enquote {\bibinfo {title} {Numerical evaluation and
  robustness of the quantum mean-force gibbs state},}\ }\href {\doibase
  10.1103/physreva.106.012204} {\bibfield  {journal} {\bibinfo  {journal}
  {Physical Review A}\ }\textbf {\bibinfo {volume} {106}} (\bibinfo {year}
  {2022}),\ 10.1103/physreva.106.012204}\BibitemShut {NoStop}%
\bibitem [{\citenamefont {von Neumann}\ and\ \citenamefont
  {Beyer}(1955)}]{vonneumann_55}%
  \BibitemOpen
  \bibfield  {author} {\bibinfo {author} {\bibfnamefont {J.}~\bibnamefont {von
  Neumann}}\ and\ \bibinfo {author} {\bibfnamefont {R.}~\bibnamefont {Beyer}},\
  }\href {https://books.google.com/books?id=JLyCo3RO4qUC} {\emph {\bibinfo
  {title} {Mathematical Foundations of Quantum Mechanics}}},\ Goldstine Printed
  Materials\ (\bibinfo  {publisher} {Princeton University Press},\ \bibinfo
  {year} {1955})\BibitemShut {NoStop}%
\bibitem [{\citenamefont {Schr\"{o}dinger}(1927)}]{schrodinger_27}%
  \BibitemOpen
  \bibfield  {author} {\bibinfo {author} {\bibfnamefont {E.}~\bibnamefont
  {Schr\"{o}dinger}},\ }\bibfield  {title} {\enquote {\bibinfo {title}
  {Energieaustausch nach der wellenmechanik},}\ }\href {\doibase
  10.1002/andp.19273881504} {\bibfield  {journal} {\bibinfo  {journal} {Annalen
  der Physik}\ }\textbf {\bibinfo {volume} {388}},\ \bibinfo {pages} {956--968}
  (\bibinfo {year} {1927})}\BibitemShut {NoStop}%
\bibitem [{\citenamefont {von Neumann}(1929)}]{vonneumann_29}%
  \BibitemOpen
  \bibfield  {author} {\bibinfo {author} {\bibfnamefont {J.}~\bibnamefont {von
  Neumann}},\ }\bibfield  {title} {\enquote {\bibinfo {title} {Beweis des
  ergodensatzes und {des$H$}-theorems in der neuen mechanik},}\ }\href
  {\doibase 10.1007/bf01339852} {\bibfield  {journal} {\bibinfo  {journal}
  {Zeitschrift f\"{u}r Physik}\ }\textbf {\bibinfo {volume} {57}},\ \bibinfo
  {pages} {30--70} (\bibinfo {year} {1929})},\ \bibinfo {note} {english
  translation \url{https://arxiv.org/abs/1003.2133}}\BibitemShut {NoStop}%
\bibitem [{\citenamefont {Goldstein}\ \emph {et~al.}(2010)\citenamefont
  {Goldstein}, \citenamefont {Lebowitz}, \citenamefont {Mastrodonato},
  \citenamefont {Tumulka},\ and\ \citenamefont
  {Zangh{\`{\i}}}}]{goldstein_lebowitz_mastrodonato_tumulka_zanghi_10}%
  \BibitemOpen
  \bibfield  {author} {\bibinfo {author} {\bibfnamefont {S.}~\bibnamefont
  {Goldstein}}, \bibinfo {author} {\bibfnamefont {J.~L.}\ \bibnamefont
  {Lebowitz}}, \bibinfo {author} {\bibfnamefont {C.}~\bibnamefont
  {Mastrodonato}}, \bibinfo {author} {\bibfnamefont {R.}~\bibnamefont
  {Tumulka}}, \ and\ \bibinfo {author} {\bibfnamefont {N.}~\bibnamefont
  {Zangh{\`{\i}}}},\ }\bibfield  {title} {\enquote {\bibinfo {title} {Normal
  typicality and von neumann's quantum ergodic theorem},}\ }\href {\doibase
  10.1098/rspa.2009.0635} {\bibfield  {journal} {\bibinfo  {journal}
  {Proceedings of the Royal Society A: Mathematical, Physical and Engineering
  Sciences}\ }\textbf {\bibinfo {volume} {466}},\ \bibinfo {pages} {3203--3224}
  (\bibinfo {year} {2010})}\BibitemShut {NoStop}%
\bibitem [{\citenamefont {Reimann}(2007)}]{reimann_07}%
  \BibitemOpen
  \bibfield  {author} {\bibinfo {author} {\bibfnamefont {P.}~\bibnamefont
  {Reimann}},\ }\bibfield  {title} {\enquote {\bibinfo {title} {Typicality for
  generalized microcanonical ensembles},}\ }\href {\doibase
  10.1103/physrevlett.99.160404} {\bibfield  {journal} {\bibinfo  {journal}
  {Physical Review Letters}\ }\textbf {\bibinfo {volume} {99}} (\bibinfo {year}
  {2007}),\ 10.1103/physrevlett.99.160404}\BibitemShut {NoStop}%
\bibitem [{\citenamefont {Bartsch}\ and\ \citenamefont
  {Gemmer}(2009)}]{bartsch2009dynamical}%
  \BibitemOpen
  \bibfield  {author} {\bibinfo {author} {\bibfnamefont {C.}~\bibnamefont
  {Bartsch}}\ and\ \bibinfo {author} {\bibfnamefont {J.}~\bibnamefont
  {Gemmer}},\ }\bibfield  {title} {\enquote {\bibinfo {title} {Dynamical
  typicality of quantum expectation values},}\ }\href {\doibase
  10.1103/PhysRevLett.102.110403} {\bibfield  {journal} {\bibinfo  {journal}
  {Physical review letters}\ }\textbf {\bibinfo {volume} {102}},\ \bibinfo
  {pages} {110403} (\bibinfo {year} {2009})}\BibitemShut {NoStop}%
\bibitem [{\citenamefont {Sugiura}\ and\ \citenamefont
  {Shimizu}(2012)}]{sugiura_shimizu_12}%
  \BibitemOpen
  \bibfield  {author} {\bibinfo {author} {\bibfnamefont {S.}~\bibnamefont
  {Sugiura}}\ and\ \bibinfo {author} {\bibfnamefont {A.}~\bibnamefont
  {Shimizu}},\ }\bibfield  {title} {\enquote {\bibinfo {title} {Thermal pure
  quantum states at finite temperature},}\ }\href {\doibase
  10.1103/physrevlett.108.240401} {\bibfield  {journal} {\bibinfo  {journal}
  {Physical Review Letters}\ }\textbf {\bibinfo {volume} {108}} (\bibinfo
  {year} {2012}),\ 10.1103/physrevlett.108.240401}\BibitemShut {NoStop}%
\bibitem [{Note1()}]{Note1}%
  \BibitemOpen
  \bibinfo {note} {While ${H}_{\protect \textup {t}}$ may be sparse, the matrix
  exponential is not, in general, sparse.}\BibitemShut {Stop}%
\bibitem [{\citenamefont {Skilling}(1989)}]{skilling_89}%
  \BibitemOpen
  \bibfield  {author} {\bibinfo {author} {\bibfnamefont {J.}~\bibnamefont
  {Skilling}},\ }\bibfield  {title} {\enquote {\bibinfo {title} {The
  eigenvalues of mega-dimensional matrices},}\ }in\ \href {\doibase
  10.1007/978-94-015-7860-8_48} {\emph {\bibinfo {booktitle} {Maximum Entropy
  and Bayesian Methods}}}\ (\bibinfo  {publisher} {Springer Netherlands},\
  \bibinfo {year} {1989})\ pp.\ \bibinfo {pages} {455--466}\BibitemShut
  {NoStop}%
\bibitem [{\citenamefont {Jakli{\v{c}}}\ and\ \citenamefont
  {Prelov{\v{s}}ek}(1994)}]{jaklic_prelovsek_94}%
  \BibitemOpen
  \bibfield  {author} {\bibinfo {author} {\bibfnamefont {J.}~\bibnamefont
  {Jakli{\v{c}}}}\ and\ \bibinfo {author} {\bibfnamefont {P.}~\bibnamefont
  {Prelov{\v{s}}ek}},\ }\bibfield  {title} {\enquote {\bibinfo {title}
  {{L}anczos method for the calculation of finite-temperature quantities in
  correlated systems},}\ }\href {\doibase 10.1103/physrevb.49.5065} {\bibfield
  {journal} {\bibinfo  {journal} {Physical Review B}\ }\textbf {\bibinfo
  {volume} {49}},\ \bibinfo {pages} {5065--5068} (\bibinfo {year}
  {1994})}\BibitemShut {NoStop}%
\bibitem [{\citenamefont {Schnalle}\ and\ \citenamefont
  {Schnack}(2010)}]{schnalle_schnack_10}%
  \BibitemOpen
  \bibfield  {author} {\bibinfo {author} {\bibfnamefont {R.}~\bibnamefont
  {Schnalle}}\ and\ \bibinfo {author} {\bibfnamefont {J.}~\bibnamefont
  {Schnack}},\ }\bibfield  {title} {\enquote {\bibinfo {title} {Calculating the
  energy spectra of magnetic molecules: application of real- and spin-space
  symmetries},}\ }\href {\doibase 10.1080/0144235x.2010.485755} {\bibfield
  {journal} {\bibinfo  {journal} {International Reviews in Physical Chemistry}\
  }\textbf {\bibinfo {volume} {29}},\ \bibinfo {pages} {403--452} (\bibinfo
  {year} {2010})}\BibitemShut {NoStop}%
\bibitem [{\citenamefont {Wei{\ss}e}\ \emph {et~al.}(2006)\citenamefont
  {Wei{\ss}e}, \citenamefont {Wellein}, \citenamefont {Alvermann},\ and\
  \citenamefont {Fehske}}]{kpm_review_06}%
  \BibitemOpen
  \bibfield  {author} {\bibinfo {author} {\bibfnamefont {A.}~\bibnamefont
  {Wei{\ss}e}}, \bibinfo {author} {\bibfnamefont {G.}~\bibnamefont {Wellein}},
  \bibinfo {author} {\bibfnamefont {A.}~\bibnamefont {Alvermann}}, \ and\
  \bibinfo {author} {\bibfnamefont {H.}~\bibnamefont {Fehske}},\ }\bibfield
  {title} {\enquote {\bibinfo {title} {The kernel polynomial method},}\ }\href
  {\doibase 10.1103/revmodphys.78.275} {\bibfield  {journal} {\bibinfo
  {journal} {Reviews of Modern Physics}\ }\textbf {\bibinfo {volume} {78}},\
  \bibinfo {pages} {275--306} (\bibinfo {year} {2006})}\BibitemShut {NoStop}%
\bibitem [{\citenamefont {Schnack}, \citenamefont {Richter},\ and\
  \citenamefont {Steinigeweg}(2020)}]{schnack_richter_steinigeweg_20}%
  \BibitemOpen
  \bibfield  {author} {\bibinfo {author} {\bibfnamefont {J.}~\bibnamefont
  {Schnack}}, \bibinfo {author} {\bibfnamefont {J.}~\bibnamefont {Richter}}, \
  and\ \bibinfo {author} {\bibfnamefont {R.}~\bibnamefont {Steinigeweg}},\
  }\bibfield  {title} {\enquote {\bibinfo {title} {Accuracy of the
  finite-temperature {L}anczos method compared to simple typicality-based
  estimates},}\ }\href {\doibase 10.1103/physrevresearch.2.013186} {\bibfield
  {journal} {\bibinfo  {journal} {Physical Review Research}\ }\textbf {\bibinfo
  {volume} {2}} (\bibinfo {year} {2020}),\
  10.1103/physrevresearch.2.013186}\BibitemShut {NoStop}%
\bibitem [{\citenamefont {Schl\"{u}ter}\ \emph {et~al.}(2021)\citenamefont
  {Schl\"{u}ter}, \citenamefont {Gayk}, \citenamefont {Schmidt}, \citenamefont
  {Honecker},\ and\ \citenamefont
  {Schnack}}]{schulter_gayk_schmidt_honecker_schnack_21}%
  \BibitemOpen
  \bibfield  {author} {\bibinfo {author} {\bibfnamefont {H.}~\bibnamefont
  {Schl\"{u}ter}}, \bibinfo {author} {\bibfnamefont {F.}~\bibnamefont {Gayk}},
  \bibinfo {author} {\bibfnamefont {H.-J.}\ \bibnamefont {Schmidt}}, \bibinfo
  {author} {\bibfnamefont {A.}~\bibnamefont {Honecker}}, \ and\ \bibinfo
  {author} {\bibfnamefont {J.}~\bibnamefont {Schnack}},\ }\bibfield  {title}
  {\enquote {\bibinfo {title} {Accuracy of the typicality approach using
  {C}hebyshev polynomials},}\ }\href {\doibase 10.1515/zna-2021-0116}
  {\bibfield  {journal} {\bibinfo  {journal} {Zeitschrift f\"{u}r
  Naturforschung A}\ }\textbf {\bibinfo {volume} {76}},\ \bibinfo {pages}
  {823--834} (\bibinfo {year} {2021})}\BibitemShut {NoStop}%
\bibitem [{\citenamefont {Jin}\ \emph {et~al.}(2021)\citenamefont {Jin},
  \citenamefont {Willsch}, \citenamefont {Willsch}, \citenamefont {Lagemann},
  \citenamefont {Michielsen},\ and\ \citenamefont
  {De~Raedt}}]{jin_willsch_willsch_lagemann_michielsen_deraedt_21}%
  \BibitemOpen
  \bibfield  {author} {\bibinfo {author} {\bibfnamefont {F.}~\bibnamefont
  {Jin}}, \bibinfo {author} {\bibfnamefont {D.}~\bibnamefont {Willsch}},
  \bibinfo {author} {\bibfnamefont {M.}~\bibnamefont {Willsch}}, \bibinfo
  {author} {\bibfnamefont {H.}~\bibnamefont {Lagemann}}, \bibinfo {author}
  {\bibfnamefont {K.}~\bibnamefont {Michielsen}}, \ and\ \bibinfo {author}
  {\bibfnamefont {H.}~\bibnamefont {De~Raedt}},\ }\bibfield  {title} {\enquote
  {\bibinfo {title} {Random state technology},}\ }\href {\doibase
  10.7566/jpsj.90.012001} {\bibfield  {journal} {\bibinfo  {journal} {Journal
  of the Physical Society of Japan}\ }\textbf {\bibinfo {volume} {90}},\
  \bibinfo {pages} {012001} (\bibinfo {year} {2021})}\BibitemShut {NoStop}%
\bibitem [{\citenamefont {Han}\ \emph {et~al.}(2017)\citenamefont {Han},
  \citenamefont {Malioutov}, \citenamefont {Avron},\ and\ \citenamefont
  {Shin}}]{han_malioutov_avron_shin_17}%
  \BibitemOpen
  \bibfield  {author} {\bibinfo {author} {\bibfnamefont {I.}~\bibnamefont
  {Han}}, \bibinfo {author} {\bibfnamefont {D.}~\bibnamefont {Malioutov}},
  \bibinfo {author} {\bibfnamefont {H.}~\bibnamefont {Avron}}, \ and\ \bibinfo
  {author} {\bibfnamefont {J.}~\bibnamefont {Shin}},\ }\bibfield  {title}
  {\enquote {\bibinfo {title} {Approximating spectral sums of large-scale
  matrices using stochastic {C}hebyshev approximations},}\ }\href {\doibase
  10.1137/16m1078148} {\bibfield  {journal} {\bibinfo  {journal} {{SIAM}
  Journal on Scientific Computing}\ }\textbf {\bibinfo {volume} {39}},\
  \bibinfo {pages} {A1558--A1585} (\bibinfo {year} {2017})}\BibitemShut
  {NoStop}%
\bibitem [{\citenamefont {Ubaru}, \citenamefont {Chen},\ and\ \citenamefont
  {Saad}(2017)}]{ubaru_chen_saad_17}%
  \BibitemOpen
  \bibfield  {author} {\bibinfo {author} {\bibfnamefont {S.}~\bibnamefont
  {Ubaru}}, \bibinfo {author} {\bibfnamefont {J.}~\bibnamefont {Chen}}, \ and\
  \bibinfo {author} {\bibfnamefont {Y.}~\bibnamefont {Saad}},\ }\bibfield
  {title} {\enquote {\bibinfo {title} {Fast estimation of {$\tr(f(A))$} via
  stochastic {L}anczos quadrature},}\ }\href {\doibase 10.1137/16m1104974}
  {\bibfield  {journal} {\bibinfo  {journal} {{SIAM} Journal on Matrix Analysis
  and Applications}\ }\textbf {\bibinfo {volume} {38}},\ \bibinfo {pages}
  {1075--1099} (\bibinfo {year} {2017})}\BibitemShut {NoStop}%
\bibitem [{\citenamefont {Chen}, \citenamefont {Trogdon},\ and\ \citenamefont
  {Ubaru}(2022)}]{chen_trogdon_ubaru_22}%
  \BibitemOpen
  \bibfield  {author} {\bibinfo {author} {\bibfnamefont {T.}~\bibnamefont
  {Chen}}, \bibinfo {author} {\bibfnamefont {T.}~\bibnamefont {Trogdon}}, \
  and\ \bibinfo {author} {\bibfnamefont {S.}~\bibnamefont {Ubaru}},\
  }\href@noop {} {\enquote {\bibinfo {title} {Randomized matrix-free quadrature
  for spectrum and spectral sum approximation},}\ } (\bibinfo {year} {2022}),\
  \Eprint {http://arxiv.org/abs/2204.01941} {arXiv:2204.01941 [math.NA]}
  \BibitemShut {NoStop}%
\bibitem [{\citenamefont {Maziero}(2017)}]{maziero_17}%
  \BibitemOpen
  \bibfield  {author} {\bibinfo {author} {\bibfnamefont {J.}~\bibnamefont
  {Maziero}},\ }\bibfield  {title} {\enquote {\bibinfo {title} {Computing
  partial traces and reduced density matrices},}\ }\href {\doibase
  10.1142/s012918311750005x} {\bibfield  {journal} {\bibinfo  {journal}
  {International Journal of Modern Physics C}\ }\textbf {\bibinfo {volume}
  {28}},\ \bibinfo {pages} {1750005} (\bibinfo {year} {2017})}\BibitemShut
  {NoStop}%
\bibitem [{\citenamefont {Bai}, \citenamefont {Fahey},\ and\ \citenamefont
  {Golub}(1996)}]{bai_fahey_golub_96}%
  \BibitemOpen
  \bibfield  {author} {\bibinfo {author} {\bibfnamefont {Z.}~\bibnamefont
  {Bai}}, \bibinfo {author} {\bibfnamefont {G.}~\bibnamefont {Fahey}}, \ and\
  \bibinfo {author} {\bibfnamefont {G.}~\bibnamefont {Golub}},\ }\bibfield
  {title} {\enquote {\bibinfo {title} {Some large-scale matrix computation
  problems},}\ }\href {\doibase 10.1016/0377-0427(96)00018-0} {\bibfield
  {journal} {\bibinfo  {journal} {Journal of Computational and Applied
  Mathematics}\ }\textbf {\bibinfo {volume} {74}},\ \bibinfo {pages} {71--89}
  (\bibinfo {year} {1996})}\BibitemShut {NoStop}%
\bibitem [{\citenamefont {Gibbs}(1902)}]{gibbs1902elementary}%
  \BibitemOpen
  \bibfield  {author} {\bibinfo {author} {\bibfnamefont {J.}~\bibnamefont
  {Gibbs}},\ }\href@noop {} {\emph {\bibinfo {title} {Elementary Principles in
  Statistical Mechanics: Developed with Especial Reference to The Rational
  Foundation of Thermodynamics}}}\ (\bibinfo  {publisher} {C. Scribner's
  sons},\ \bibinfo {year} {1902})\BibitemShut {NoStop}%
\bibitem [{\citenamefont {Lu}\ and\ \citenamefont
  {Qian}(2022)}]{lu2022emergence}%
  \BibitemOpen
  \bibfield  {author} {\bibinfo {author} {\bibfnamefont {Z.}~\bibnamefont
  {Lu}}\ and\ \bibinfo {author} {\bibfnamefont {H.}~\bibnamefont {Qian}},\
  }\bibfield  {title} {\enquote {\bibinfo {title} {Emergence and breaking of
  duality symmetry in generalized fundamental thermodynamic relations},}\
  }\href {\doibase 10.1103/PhysRevLett.128.150603} {\bibfield  {journal}
  {\bibinfo  {journal} {Phys. Rev. Lett.}\ }\textbf {\bibinfo {volume} {128}},\
  \bibinfo {pages} {150603} (\bibinfo {year} {2022})}\BibitemShut {NoStop}%
\bibitem [{\citenamefont {Golub}\ and\ \citenamefont
  {Meurant}(2009)}]{golub_meurant_09}%
  \BibitemOpen
  \bibfield  {author} {\bibinfo {author} {\bibfnamefont {G.}~\bibnamefont
  {Golub}}\ and\ \bibinfo {author} {\bibfnamefont {G.}~\bibnamefont
  {Meurant}},\ }\href@noop {} {\emph {\bibinfo {title} {Matrices, Moments and
  Quadrature with Applications}}},\ Princeton Series in Applied Mathematics\
  (\bibinfo  {publisher} {Princeton University Press},\ \bibinfo {year}
  {2009})\BibitemShut {NoStop}%
\bibitem [{\citenamefont {Avron}\ and\ \citenamefont
  {Toledo}(2011)}]{avron_toledo_11}%
  \BibitemOpen
  \bibfield  {author} {\bibinfo {author} {\bibfnamefont {H.}~\bibnamefont
  {Avron}}\ and\ \bibinfo {author} {\bibfnamefont {S.}~\bibnamefont {Toledo}},\
  }\bibfield  {title} {\enquote {\bibinfo {title} {Randomized algorithms for
  estimating the trace of an implicit symmetric positive semi-definite
  matrix},}\ }\href {\doibase 10.1145/1944345.1944349} {\bibfield  {journal}
  {\bibinfo  {journal} {Journal of the {ACM}}\ }\textbf {\bibinfo {volume}
  {58}},\ \bibinfo {pages} {1--34} (\bibinfo {year} {2011})}\BibitemShut
  {NoStop}%
\bibitem [{\citenamefont {Roosta-Khorasani}\ and\ \citenamefont
  {Ascher}(2014)}]{roostakhorasani_ascher_14}%
  \BibitemOpen
  \bibfield  {author} {\bibinfo {author} {\bibfnamefont {F.}~\bibnamefont
  {Roosta-Khorasani}}\ and\ \bibinfo {author} {\bibfnamefont {U.}~\bibnamefont
  {Ascher}},\ }\bibfield  {title} {\enquote {\bibinfo {title} {Improved bounds
  on sample size for implicit matrix trace estimators},}\ }\href {\doibase
  10.1007/s10208-014-9220-1} {\bibfield  {journal} {\bibinfo  {journal}
  {Foundations of Computational Mathematics}\ }\textbf {\bibinfo {volume}
  {15}},\ \bibinfo {pages} {1187--1212} (\bibinfo {year} {2014})}\BibitemShut
  {NoStop}%
\bibitem [{\citenamefont {Cortinovis}\ and\ \citenamefont
  {Kressner}(2021)}]{cortinovis_kressner_21}%
  \BibitemOpen
  \bibfield  {author} {\bibinfo {author} {\bibfnamefont {A.}~\bibnamefont
  {Cortinovis}}\ and\ \bibinfo {author} {\bibfnamefont {D.}~\bibnamefont
  {Kressner}},\ }\bibfield  {title} {\enquote {\bibinfo {title} {On randomized
  trace estimates for indefinite matrices with an application to
  determinants},}\ }\href {\doibase 10.1007/s10208-021-09525-9} {\bibfield
  {journal} {\bibinfo  {journal} {Foundations of Computational Mathematics}\ }
  (\bibinfo {year} {2021}),\ 10.1007/s10208-021-09525-9}\BibitemShut {NoStop}%
\bibitem [{\citenamefont {Meyer}\ \emph {et~al.}(2021)\citenamefont {Meyer},
  \citenamefont {Musco}, \citenamefont {Musco},\ and\ \citenamefont
  {Woodruff}}]{meyer_musco_musco_woodruff_21}%
  \BibitemOpen
  \bibfield  {author} {\bibinfo {author} {\bibfnamefont {R.~A.}\ \bibnamefont
  {Meyer}}, \bibinfo {author} {\bibfnamefont {C.}~\bibnamefont {Musco}},
  \bibinfo {author} {\bibfnamefont {C.}~\bibnamefont {Musco}}, \ and\ \bibinfo
  {author} {\bibfnamefont {D.~P.}\ \bibnamefont {Woodruff}},\ }\bibfield
  {title} {\enquote {\bibinfo {title} {Hutch++: Optimal stochastic trace
  estimation},}\ }in\ \href {\doibase 10.1137/1.9781611976496.16} {\emph
  {\bibinfo {booktitle} {Symposium on Simplicity in Algorithms ({SOSA})}}}\
  (\bibinfo  {publisher} {Society for Industrial and Applied Mathematics},\
  \bibinfo {year} {2021})\ pp.\ \bibinfo {pages} {142--155}\BibitemShut
  {NoStop}%
\bibitem [{\citenamefont {Chen}, \citenamefont {Trogdon},\ and\ \citenamefont
  {Ubaru}(2021)}]{chen_trogdon_ubaru_21}%
  \BibitemOpen
  \bibfield  {author} {\bibinfo {author} {\bibfnamefont {T.}~\bibnamefont
  {Chen}}, \bibinfo {author} {\bibfnamefont {T.}~\bibnamefont {Trogdon}}, \
  and\ \bibinfo {author} {\bibfnamefont {S.}~\bibnamefont {Ubaru}},\ }\bibfield
   {title} {\enquote {\bibinfo {title} {Analysis of stochastic {L}anczos
  quadrature for spectrum approximation},}\ }in\ \href@noop {} {\emph {\bibinfo
  {booktitle} {Proceedings of the 37th International Conference on Machine
  Learning}}},\ \bibinfo {series and number} {Proceedings of Machine Learning
  Research}\ (\bibinfo  {publisher} {PMLR},\ \bibinfo {year} {2021})\ \Eprint
  {http://arxiv.org/abs/2105.06595} {arXiv:2105.06595 [cs.DS]} \BibitemShut
  {NoStop}%
\bibitem [{\citenamefont {Persson}, \citenamefont {Cortinovis},\ and\
  \citenamefont {Kressner}(2022)}]{persson_cortinovis_kressner_22}%
  \BibitemOpen
  \bibfield  {author} {\bibinfo {author} {\bibfnamefont {D.}~\bibnamefont
  {Persson}}, \bibinfo {author} {\bibfnamefont {A.}~\bibnamefont {Cortinovis}},
  \ and\ \bibinfo {author} {\bibfnamefont {D.}~\bibnamefont {Kressner}},\
  }\bibfield  {title} {\enquote {\bibinfo {title} {Improved variants of the
  hutch++ algorithm for trace estimation},}\ }\href {\doibase
  10.1137/21m1447623} {\bibfield  {journal} {\bibinfo  {journal} {{SIAM}
  Journal on Matrix Analysis and Applications}\ }\textbf {\bibinfo {volume}
  {43}},\ \bibinfo {pages} {1162--1185} (\bibinfo {year} {2022})}\BibitemShut
  {NoStop}%
\bibitem [{\citenamefont {Lin}(2016)}]{lin_16}%
  \BibitemOpen
  \bibfield  {author} {\bibinfo {author} {\bibfnamefont {L.}~\bibnamefont
  {Lin}},\ }\bibfield  {title} {\enquote {\bibinfo {title} {Randomized
  estimation of spectral densities of large matrices made accurate},}\ }\href
  {\doibase 10.1007/s00211-016-0837-7} {\bibfield  {journal} {\bibinfo
  {journal} {Numerische Mathematik}\ }\textbf {\bibinfo {volume} {136}},\
  \bibinfo {pages} {183--213} (\bibinfo {year} {2016})}\BibitemShut {NoStop}%
\bibitem [{\citenamefont {Gambhir}, \citenamefont {Stathopoulos},\ and\
  \citenamefont {Orginos}(2017)}]{gambhir_stathopoulos_orginos_17}%
  \BibitemOpen
  \bibfield  {author} {\bibinfo {author} {\bibfnamefont {A.~S.}\ \bibnamefont
  {Gambhir}}, \bibinfo {author} {\bibfnamefont {A.}~\bibnamefont
  {Stathopoulos}}, \ and\ \bibinfo {author} {\bibfnamefont {K.}~\bibnamefont
  {Orginos}},\ }\bibfield  {title} {\enquote {\bibinfo {title} {Deflation as a
  method of variance reduction for estimating the trace of a matrix inverse},}\
  }\href {\doibase 10.1137/16m1066361} {\bibfield  {journal} {\bibinfo
  {journal} {{SIAM} Journal on Scientific Computing}\ }\textbf {\bibinfo
  {volume} {39}},\ \bibinfo {pages} {A532--A558} (\bibinfo {year}
  {2017})}\BibitemShut {NoStop}%
\bibitem [{\citenamefont {Chen}\ and\ \citenamefont
  {Hallman}(2022)}]{chen_hallman_22}%
  \BibitemOpen
  \bibfield  {author} {\bibinfo {author} {\bibfnamefont {T.}~\bibnamefont
  {Chen}}\ and\ \bibinfo {author} {\bibfnamefont {E.}~\bibnamefont {Hallman}},\
  }\href@noop {} {\enquote {\bibinfo {title} {Krylov-aware stochastic trace
  estimation},}\ } (\bibinfo {year} {2022}),\ \Eprint
  {http://arxiv.org/abs/2205.01736} {arXiv:2205.01736 [math.NA]} \BibitemShut
  {NoStop}%
\bibitem [{\citenamefont {Trefethen}(2019)}]{trefethen_19}%
  \BibitemOpen
  \bibfield  {author} {\bibinfo {author} {\bibfnamefont {L.~N.}\ \bibnamefont
  {Trefethen}},\ }\href@noop {} {\emph {\bibinfo {title} {Approximation Theory
  and Approximation Practice, Extended Edition}}}\ (\bibinfo  {publisher}
  {SIAM},\ \bibinfo {year} {2019})\BibitemShut {NoStop}%
\bibitem [{\citenamefont {Aichhorn}\ \emph {et~al.}(2003)\citenamefont
  {Aichhorn}, \citenamefont {Daghofer}, \citenamefont {Evertz},\ and\
  \citenamefont {von~der Linden}}]{aichhorn_daghofer_evertz_vondelinden_03}%
  \BibitemOpen
  \bibfield  {author} {\bibinfo {author} {\bibfnamefont {M.}~\bibnamefont
  {Aichhorn}}, \bibinfo {author} {\bibfnamefont {M.}~\bibnamefont {Daghofer}},
  \bibinfo {author} {\bibfnamefont {H.~G.}\ \bibnamefont {Evertz}}, \ and\
  \bibinfo {author} {\bibfnamefont {W.}~\bibnamefont {von~der Linden}},\
  }\bibfield  {title} {\enquote {\bibinfo {title} {Low-temperature {L}anczos
  method for strongly correlated systems},}\ }\href {\doibase
  10.1103/physrevb.67.161103} {\bibfield  {journal} {\bibinfo  {journal}
  {Physical Review B}\ }\textbf {\bibinfo {volume} {67}} (\bibinfo {year}
  {2003}),\ 10.1103/physrevb.67.161103}\BibitemShut {NoStop}%
\bibitem [{\citenamefont {Paige}(1976)}]{paige_76}%
  \BibitemOpen
  \bibfield  {author} {\bibinfo {author} {\bibfnamefont {C.~C.}\ \bibnamefont
  {Paige}},\ }\bibfield  {title} {\enquote {\bibinfo {title} {{Error Analysis
  of the {L}anczos Algorithm for Tridiagonalizing a Symmetric Matrix}},}\
  }\href {\doibase 10.1093/imamat/18.3.341} {\bibfield  {journal} {\bibinfo
  {journal} {IMA Journal of Applied Mathematics}\ }\textbf {\bibinfo {volume}
  {18}},\ \bibinfo {pages} {341--349} (\bibinfo {year} {1976})}\BibitemShut
  {NoStop}%
\bibitem [{\citenamefont {Paige}(1980)}]{paige_80}%
  \BibitemOpen
  \bibfield  {author} {\bibinfo {author} {\bibfnamefont {C.~C.}\ \bibnamefont
  {Paige}},\ }\bibfield  {title} {\enquote {\bibinfo {title} {Accuracy and
  effectiveness of the {L}anczos algorithm for the symmetric eigenproblem},}\
  }\href {\doibase 10.1016/0024-3795(80)90167-6} {\bibfield  {journal}
  {\bibinfo  {journal} {Linear Algebra and its Applications}\ }\textbf
  {\bibinfo {volume} {34}},\ \bibinfo {pages} {235 -- 258} (\bibinfo {year}
  {1980})}\BibitemShut {NoStop}%
\bibitem [{\citenamefont {Musco}, \citenamefont {Musco},\ and\ \citenamefont
  {Sidford}(2018)}]{musco_musco_sidford_18}%
  \BibitemOpen
  \bibfield  {author} {\bibinfo {author} {\bibfnamefont {C.}~\bibnamefont
  {Musco}}, \bibinfo {author} {\bibfnamefont {C.}~\bibnamefont {Musco}}, \ and\
  \bibinfo {author} {\bibfnamefont {A.}~\bibnamefont {Sidford}},\ }\bibfield
  {title} {\enquote {\bibinfo {title} {Stability of the {L}anczos method for
  matrix function approximation},}\ }in\ \href@noop {} {\emph {\bibinfo
  {booktitle} {Proceedings of the Twenty-Ninth Annual ACM-SIAM Symposium on
  Discrete Algorithms}}},\ \bibinfo {series and number} {SODA ’18}\ (\bibinfo
   {publisher} {Society for Industrial and Applied Mathematics},\ \bibinfo
  {address} {USA},\ \bibinfo {year} {2018})\ p.\ \bibinfo {pages}
  {1605–1624}\BibitemShut {NoStop}%
\bibitem [{\citenamefont {Knizhnerman}(1996)}]{knizhnerman_96}%
  \BibitemOpen
  \bibfield  {author} {\bibinfo {author} {\bibfnamefont {L.~A.}\ \bibnamefont
  {Knizhnerman}},\ }\bibfield  {title} {\enquote {\bibinfo {title} {The simple
  {L}anczos procedure: Estimates of the error of the {G}auss quadrature formula
  and their applications},}\ }\href@noop {} {\bibfield  {journal} {\bibinfo
  {journal} {Comput. Math. Math. Phys.}\ }\textbf {\bibinfo {volume} {36}},\
  \bibinfo {pages} {1481–1492} (\bibinfo {year} {1996})}\BibitemShut
  {NoStop}%
\bibitem [{\citenamefont {Rost}\ \emph {et~al.}(2009)\citenamefont {Rost},
  \citenamefont {Perry}, \citenamefont {Mercure}, \citenamefont {Mackenzie},\
  and\ \citenamefont {Grigera}}]{rost_perry_mercure_mackenzie_grigera_09}%
  \BibitemOpen
  \bibfield  {author} {\bibinfo {author} {\bibfnamefont {A.~W.}\ \bibnamefont
  {Rost}}, \bibinfo {author} {\bibfnamefont {R.~S.}\ \bibnamefont {Perry}},
  \bibinfo {author} {\bibfnamefont {J.-F.}\ \bibnamefont {Mercure}}, \bibinfo
  {author} {\bibfnamefont {A.~P.}\ \bibnamefont {Mackenzie}}, \ and\ \bibinfo
  {author} {\bibfnamefont {S.~A.}\ \bibnamefont {Grigera}},\ }\bibfield
  {title} {\enquote {\bibinfo {title} {Entropy landscape of phase formation
  associated with quantum criticality in
  $\textrm{Sr}_3\textrm{Ru}_2\textrm{O}_7$},}\ }\href {\doibase
  10.1126/science.1176627} {\bibfield  {journal} {\bibinfo  {journal}
  {Science}\ }\textbf {\bibinfo {volume} {325}},\ \bibinfo {pages} {1360--1363}
  (\bibinfo {year} {2009})}\BibitemShut {NoStop}%
\bibitem [{\citenamefont {Werlang}\ \emph {et~al.}(2010)\citenamefont
  {Werlang}, \citenamefont {Trippe}, \citenamefont {Ribeiro},\ and\
  \citenamefont {Rigolin}}]{werlang_troppe_ribeiro_rigolin_10}%
  \BibitemOpen
  \bibfield  {author} {\bibinfo {author} {\bibfnamefont {T.}~\bibnamefont
  {Werlang}}, \bibinfo {author} {\bibfnamefont {C.}~\bibnamefont {Trippe}},
  \bibinfo {author} {\bibfnamefont {G.~A.~P.}\ \bibnamefont {Ribeiro}}, \ and\
  \bibinfo {author} {\bibfnamefont {G.}~\bibnamefont {Rigolin}},\ }\bibfield
  {title} {\enquote {\bibinfo {title} {Quantum correlations in spin chains at
  finite temperatures and quantum phase transitions},}\ }\href {\doibase
  10.1103/physrevlett.105.095702} {\bibfield  {journal} {\bibinfo  {journal}
  {Physical Review Letters}\ }\textbf {\bibinfo {volume} {105}} (\bibinfo
  {year} {2010}),\ 10.1103/physrevlett.105.095702}\BibitemShut {NoStop}%
\bibitem [{\citenamefont {Koffel}, \citenamefont {Lewenstein},\ and\
  \citenamefont {Tagliacozzo}(2012)}]{koffel_lewenstein_tagliacozzo_12}%
  \BibitemOpen
  \bibfield  {author} {\bibinfo {author} {\bibfnamefont {T.}~\bibnamefont
  {Koffel}}, \bibinfo {author} {\bibfnamefont {M.}~\bibnamefont {Lewenstein}},
  \ and\ \bibinfo {author} {\bibfnamefont {L.}~\bibnamefont {Tagliacozzo}},\
  }\bibfield  {title} {\enquote {\bibinfo {title} {Entanglement entropy for the
  long-range ising chain in a transverse field},}\ }\href {\doibase
  10.1103/physrevlett.109.267203} {\bibfield  {journal} {\bibinfo  {journal}
  {Physical Review Letters}\ }\textbf {\bibinfo {volume} {109}} (\bibinfo
  {year} {2012}),\ 10.1103/physrevlett.109.267203}\BibitemShut {NoStop}%
\bibitem [{\citenamefont {Breunig}\ \emph {et~al.}(2017)\citenamefont
  {Breunig}, \citenamefont {Garst}, \citenamefont {Kl\"{u}mper}, \citenamefont
  {Rohrkamp}, \citenamefont {Turnbull},\ and\ \citenamefont
  {Lorenz}}]{breunig_garst_klumper_rohrkamp_turnbull_lorenz_17}%
  \BibitemOpen
  \bibfield  {author} {\bibinfo {author} {\bibfnamefont {O.}~\bibnamefont
  {Breunig}}, \bibinfo {author} {\bibfnamefont {M.}~\bibnamefont {Garst}},
  \bibinfo {author} {\bibfnamefont {A.}~\bibnamefont {Kl\"{u}mper}}, \bibinfo
  {author} {\bibfnamefont {J.}~\bibnamefont {Rohrkamp}}, \bibinfo {author}
  {\bibfnamefont {M.~M.}\ \bibnamefont {Turnbull}}, \ and\ \bibinfo {author}
  {\bibfnamefont {T.}~\bibnamefont {Lorenz}},\ }\bibfield  {title} {\enquote
  {\bibinfo {title} {Quantum criticality in the spin-1/2 heisenberg chain
  system copper pyrazine dinitrate},}\ }\href {\doibase 10.1126/sciadv.aao3773}
  {\bibfield  {journal} {\bibinfo  {journal} {Science Advances}\ }\textbf
  {\bibinfo {volume} {3}} (\bibinfo {year} {2017}),\
  10.1126/sciadv.aao3773}\BibitemShut {NoStop}%
\bibitem [{\citenamefont {Sondhi}\ \emph {et~al.}(1997)\citenamefont {Sondhi},
  \citenamefont {Girvin}, \citenamefont {Carini},\ and\ \citenamefont
  {Shahar}}]{sondhi_girvin_carini_97}%
  \BibitemOpen
  \bibfield  {author} {\bibinfo {author} {\bibfnamefont {S.~L.}\ \bibnamefont
  {Sondhi}}, \bibinfo {author} {\bibfnamefont {S.~M.}\ \bibnamefont {Girvin}},
  \bibinfo {author} {\bibfnamefont {J.~P.}\ \bibnamefont {Carini}}, \ and\
  \bibinfo {author} {\bibfnamefont {D.}~\bibnamefont {Shahar}},\ }\bibfield
  {title} {\enquote {\bibinfo {title} {Continuous quantum phase transitions},}\
  }\href {\doibase 10.1103/revmodphys.69.315} {\bibfield  {journal} {\bibinfo
  {journal} {Reviews of Modern Physics}\ }\textbf {\bibinfo {volume} {69}},\
  \bibinfo {pages} {315--333} (\bibinfo {year} {1997})}\BibitemShut {NoStop}%
\bibitem [{\citenamefont {Jarzynski}(2017)}]{jarzynski_17}%
  \BibitemOpen
  \bibfield  {author} {\bibinfo {author} {\bibfnamefont {C.}~\bibnamefont
  {Jarzynski}},\ }\bibfield  {title} {\enquote {\bibinfo {title} {Stochastic
  and macroscopic thermodynamics of strongly coupled systems},}\ }\href
  {\doibase 10.1103/PhysRevX.7.011008} {\bibfield  {journal} {\bibinfo
  {journal} {Physical Review X}\ }\textbf {\bibinfo {volume} {7}},\ \bibinfo
  {pages} {011008} (\bibinfo {year} {2017})}\BibitemShut {NoStop}%
\bibitem [{\citenamefont {Gelin}\ and\ \citenamefont
  {Thoss}(2009)}]{gelin_thoss_09}%
  \BibitemOpen
  \bibfield  {author} {\bibinfo {author} {\bibfnamefont {M.~F.}\ \bibnamefont
  {Gelin}}\ and\ \bibinfo {author} {\bibfnamefont {M.}~\bibnamefont {Thoss}},\
  }\bibfield  {title} {\enquote {\bibinfo {title} {Thermodynamics of a
  subensemble of a canonical ensemble},}\ }\href {\doibase
  10.1103/PhysRevE.79.051121} {\bibfield  {journal} {\bibinfo  {journal}
  {Physical Review E}\ }\textbf {\bibinfo {volume} {79}},\ \bibinfo {pages}
  {051121} (\bibinfo {year} {2009})}\BibitemShut {NoStop}%
\bibitem [{\citenamefont {Seifert}(2016)}]{seifert_16}%
  \BibitemOpen
  \bibfield  {author} {\bibinfo {author} {\bibfnamefont {U.}~\bibnamefont
  {Seifert}},\ }\bibfield  {title} {\enquote {\bibinfo {title} {First and
  second law of thermodynamics at strong coupling},}\ }\href {\doibase
  10.1103/PhysRevLett.116.020601} {\bibfield  {journal} {\bibinfo  {journal}
  {Physical review letters}\ }\textbf {\bibinfo {volume} {116}},\ \bibinfo
  {pages} {020601} (\bibinfo {year} {2016})}\BibitemShut {NoStop}%
\bibitem [{\citenamefont {Hsiang}\ and\ \citenamefont
  {Hu}(2018)}]{hsiang_hu_18}%
  \BibitemOpen
  \bibfield  {author} {\bibinfo {author} {\bibfnamefont {J.-T.}\ \bibnamefont
  {Hsiang}}\ and\ \bibinfo {author} {\bibfnamefont {B.-L.}\ \bibnamefont
  {Hu}},\ }\bibfield  {title} {\enquote {\bibinfo {title} {Quantum
  thermodynamics at strong coupling: operator thermodynamic functions and
  relations},}\ }\href {\doibase 10.3390/e20060423} {\bibfield  {journal}
  {\bibinfo  {journal} {Entropy}\ }\textbf {\bibinfo {volume} {20}},\ \bibinfo
  {pages} {423} (\bibinfo {year} {2018})}\BibitemShut {NoStop}%
\bibitem [{\citenamefont {Ben-Naim}(2013)}]{arieh_13}%
  \BibitemOpen
  \bibfield  {author} {\bibinfo {author} {\bibfnamefont {A.~Y.}\ \bibnamefont
  {Ben-Naim}},\ }\href@noop {} {\emph {\bibinfo {title} {Solvation
  thermodynamics}}}\ (\bibinfo  {publisher} {Springer Science \& Business
  Media},\ \bibinfo {year} {2013})\BibitemShut {NoStop}%
\bibitem [{\citenamefont {Gong}\ \emph {et~al.}(2020)\citenamefont {Gong},
  \citenamefont {Wang}, \citenamefont {Zhang}, \citenamefont {Qiao},
  \citenamefont {Xu}, \citenamefont {Zheng},\ and\ \citenamefont
  {Yan}}]{gong2020equilibrium}%
  \BibitemOpen
  \bibfield  {author} {\bibinfo {author} {\bibfnamefont {H.}~\bibnamefont
  {Gong}}, \bibinfo {author} {\bibfnamefont {Y.}~\bibnamefont {Wang}}, \bibinfo
  {author} {\bibfnamefont {H.-D.}\ \bibnamefont {Zhang}}, \bibinfo {author}
  {\bibfnamefont {Q.}~\bibnamefont {Qiao}}, \bibinfo {author} {\bibfnamefont
  {R.-X.}\ \bibnamefont {Xu}}, \bibinfo {author} {\bibfnamefont
  {X.}~\bibnamefont {Zheng}}, \ and\ \bibinfo {author} {\bibfnamefont
  {Y.}~\bibnamefont {Yan}},\ }\bibfield  {title} {\enquote {\bibinfo {title}
  {Equilibrium and transient thermodynamics: A unified dissipaton-space
  approach},}\ }\href {\doibase https://doi.org/10.1063/5.0021203} {\bibfield
  {journal} {\bibinfo  {journal} {The Journal of Chemical Physics}\ }\textbf
  {\bibinfo {volume} {153}},\ \bibinfo {pages} {154111} (\bibinfo {year}
  {2020})}\BibitemShut {NoStop}%
\bibitem [{\citenamefont {Long}\ \emph {et~al.}(2003)\citenamefont {Long},
  \citenamefont {Prelov{\v{s}}ek}, \citenamefont {Shawish}, \citenamefont
  {Karadamoglou},\ and\ \citenamefont
  {Zotos}}]{long_prelovsek_elshawish_karadamoglou_zotos_03}%
  \BibitemOpen
  \bibfield  {author} {\bibinfo {author} {\bibfnamefont {M.~W.}\ \bibnamefont
  {Long}}, \bibinfo {author} {\bibfnamefont {P.}~\bibnamefont
  {Prelov{\v{s}}ek}}, \bibinfo {author} {\bibfnamefont {S.~E.}\ \bibnamefont
  {Shawish}}, \bibinfo {author} {\bibfnamefont {J.}~\bibnamefont
  {Karadamoglou}}, \ and\ \bibinfo {author} {\bibfnamefont {X.}~\bibnamefont
  {Zotos}},\ }\bibfield  {title} {\enquote {\bibinfo {title}
  {Finite-temperature dynamical correlations using the microcanonical ensemble
  and the lanczos algorithm},}\ }\href {\doibase 10.1103/physrevb.68.235106}
  {\bibfield  {journal} {\bibinfo  {journal} {Physical Review B}\ }\textbf
  {\bibinfo {volume} {68}} (\bibinfo {year} {2003}),\
  10.1103/physrevb.68.235106}\BibitemShut {NoStop}%
\bibitem [{\citenamefont {Bujanovic}\ and\ \citenamefont
  {Kressner}(2021)}]{bujanovic_kressner_21}%
  \BibitemOpen
  \bibfield  {author} {\bibinfo {author} {\bibfnamefont {Z.}~\bibnamefont
  {Bujanovic}}\ and\ \bibinfo {author} {\bibfnamefont {D.}~\bibnamefont
  {Kressner}},\ }\bibfield  {title} {\enquote {\bibinfo {title} {Norm and trace
  estimation with random rank-one vectors},}\ }\href {\doibase
  10.1137/20m1331718} {\bibfield  {journal} {\bibinfo  {journal} {{SIAM}
  Journal on Matrix Analysis and Applications}\ }\textbf {\bibinfo {volume}
  {42}},\ \bibinfo {pages} {202--223} (\bibinfo {year} {2021})}\BibitemShut
  {NoStop}%
\bibitem [{\citenamefont {Vershynin}(2020)}]{vershynin_20}%
  \BibitemOpen
  \bibfield  {author} {\bibinfo {author} {\bibfnamefont {R.}~\bibnamefont
  {Vershynin}},\ }\bibfield  {title} {\enquote {\bibinfo {title} {Concentration
  inequalities for random tensors},}\ }\href {\doibase 10.3150/20-bej1218}
  {\bibfield  {journal} {\bibinfo  {journal} {Bernoulli}\ }\textbf {\bibinfo
  {volume} {26}} (\bibinfo {year} {2020}),\ 10.3150/20-bej1218}\BibitemShut
  {NoStop}%
\bibitem [{\citenamefont {Bamberger}, \citenamefont {Krahmer},\ and\
  \citenamefont {Ward}(2021)}]{bamberger_krahmer_ward_21}%
  \BibitemOpen
  \bibfield  {author} {\bibinfo {author} {\bibfnamefont {S.}~\bibnamefont
  {Bamberger}}, \bibinfo {author} {\bibfnamefont {F.}~\bibnamefont {Krahmer}},
  \ and\ \bibinfo {author} {\bibfnamefont {R.}~\bibnamefont {Ward}},\
  }\href@noop {} {\enquote {\bibinfo {title} {The hanson-wright inequality for
  random tensors},}\ } (\bibinfo {year} {2021}),\ \Eprint
  {http://arxiv.org/abs/2106.13345} {arXiv:2106.13345 [math.PR]} \BibitemShut
  {NoStop}%
\end{thebibliography}%

\end{document}